\documentclass{article}

\usepackage[preprint]{neurips_2026}

\usepackage[utf8]{inputenc}
\usepackage[T1]{fontenc}
\usepackage{hyperref}
\usepackage{url}
\usepackage{booktabs}
\usepackage{amsfonts}
\usepackage{nicefrac}
\usepackage{microtype}
\usepackage{amsmath}
\usepackage{soul}
\usepackage{amssymb}
\usepackage{enumitem}
\usepackage{graphicx}
\usepackage{multirow}
\usepackage[table]{xcolor}
\usepackage{tikz}
\usepackage{array}
\usepackage{tabularx}

\definecolor{tablegray}{gray}{0.92}
\definecolor{baselineblue}{RGB}{235,245,255}
\definecolor{occamrow}{RGB}{240,245,249}

\newcommand{\reprcell}[2]{#1\,{\fontsize{6}{6}\selectfont$\pm$\makebox[10.5pt][l]{#2}}}
\newcommand{\reprpoint}[1]{#1\phantom{\,{\fontsize{6}{6}\selectfont$\pm$\makebox[10.5pt][l]{0.00}}}}

\sethlcolor{baselineblue}

\makeatletter
\newcommand{\listofappendices}{%
  \section*{Appendix contents}%
  \@starttoc{apx}%
}
\makeatother
\title{\textsc{Occam}: Improved Single-Cell Representation Learning and Perturbation Effect Prediction}

\author{%
\normalfont
\begin{tabular}{c}
\textbf{Ihab Bendidi}$^{1,3,}$\thanks{Correspondence: \href{mailto:ihab@rivercell.ai}{ihab@rivercell.ai}.}\kern3pt$^{,}$\thanks{Work done at Valence Labs.} \quad
\textbf{Kian Kenyon-Dean}$^{2}$ \quad
\textbf{Alisandra K. Denton}$^{2}$ \\
\textbf{Auguste Genovesio}$^{3,}$\thanks{Equal advising.} \quad
\textbf{Emmanuel Noutahi}$^{2,\dagger,\ddagger}$ \\
\\[-0.5em]
$^{1}$Rivercell, Paris, France \\
$^{2}$Valence Labs, Montreal, Canada \\
$^{3}$\'{E}cole Normale Sup\'{e}rieure, PSL University, Paris, France
\end{tabular}%
}
\begin{document}
\raggedbottom

\maketitle

\begin{abstract}
  Predicting the effects of genetic perturbations requires models that generalize beyond observed cell contexts and gene combinations. We introduce \textsc{Occam}, a multilayer perceptron (MLP) framework for representation learning and perturbation-effect prediction. Occam combines cell representations learned through denoising with gene interaction, text, and protein-sequence features. It achieves the highest aggregate representation scores for perturbation information and cell-type preservation with batch integration. In cross-context few-shot prediction, a perturbation and a cell context are each observed during training, but never together. Occam better captures the measured response patterns for these unobserved combinations than four retrained competitors or direct transfer of the same perturbation's measured effects from other training cell lines. Without fine-tuning, Occam best distinguishes perturbations among the evaluated models and simple baselines in cell lines held out from predictor training, across seen and unseen assays. Predicting expression levels accurately in these settings remains challenging. For double perturbations, which target two genes simultaneously, Occam composes learned single-gene responses to better capture the combined response and reduce expression error compared with competing models and summed single-gene effects. These gains extend to both strongly additive and weakly additive responses. Together, these results support simple MLPs for cell representation learning and perturbation-effect prediction across contexts and gene combinations.\footnote{Code and data are available at \url{https://github.com/IhabBendidi/Occam}.}

\end{abstract}

\section{Introduction}
Predicting how cells respond to genetic perturbations could help prioritize experiments and characterize gene function. Single-cell perturbation screens measure RNA expression after targeted gene perturbations across many cells~\citep{replogle,nadig}. Yet the space of possible experiments grows with the number of genes, cell types, and perturbation combinations. A useful predictor must therefore generalize to conditions not measured during training: a perturbation in another cell line, or a double perturbation that targets two genes simultaneously in the same cell.

An encoder compresses an expression profile into features that support tasks such as identifying cell types or comparing perturbations. Predicting a response additionally requires learning how a specified perturbation changes that state. This relationship must be learned from separate control and perturbed populations, because RNA sequencing destroys the cell and prevents a before--after measurement of the same cell. We use \emph{perturbation-effect prediction} for the task of mapping a control expression profile and a perturbation to a predicted perturbed profile.

Transformer models provide reusable cell representations~\citep{scgpt}. However, pretrained embeddings and complex predictors do not consistently outperform simple alternatives for perturbation-effect prediction~\citep{simplebaselines1,simplebaselines2}. We therefore examine a multilayer perceptron (MLP) that processes a fixed gene-expression vector and learns relationships among genes through reconstruction.

\textsc{Occam} combines representation learning and perturbation-effect prediction in two stages. First, swap-in denoising reconstructs profiles after replacing selected gene values with measurements from another cell. Second, a predictor keeps the encoder frozen and uses gene interaction, text, and protein-sequence features to condition the encoded state on a perturbation. Its training objective supervises expression levels and the direction of change relative to controls. To predict double perturbations, we apply the same predictor successively to each target gene and average predictions from both orders.

Across our benchmarks, Occam achieves the strongest overall representation performance for retaining perturbation information and integrating cells across batches while preserving cell types. In few-shot perturbation-effect prediction, a perturbation and a cell line are each seen during training, but never together. Occam exceeds experimental reproducibility in mean agreement with reference delta profiles, which describe expression changes relative to controls. For double perturbations, Occam improves agreement with reference delta profiles and prediction of mean expression levels over competing models and the additive baseline, which sums the measured effects of the two single-gene perturbations. In cell lines held out from predictor training, Occam distinguishes perturbations better than the evaluated models and simple baselines, across both seen and unseen assays, while expression-level accuracy remains limited.

\section{Related Work}
\label{sec:related-work}
\paragraph{Cell representations.}
Transformer models, including scGPT, TranscriptFormer, STATE, AIDO.Cell, and TxFM, learn reusable cell features from expression data~\citep{scgpt,transcriptformer,state,aidocell,txfm}. Multilayer perceptron (MLP) approaches offer another route: scVI uses a variational model of expression, while scMAE reconstructs corrupted profiles and predicts which entries were replaced~\citep{scvi,scmae}. Occam builds on this denoising approach: replaced values come from a single donor cell, and reconstruction is supervised in normalized count space. We compare these nonlinear encoders with principal component analysis (PCA), ridge-regularized PCA, and a linear autoencoder. These baselines test whether nonlinear features improve on linear representations, complementing studies of simple representations and predictors~\citep{bendidi2024benchmarkpca,simplebaselines1,simplebaselines2}.

\paragraph{Perturbation-effect prediction.}
Perturbation-effect predictors estimate the expression response to a specified intervention. TxPert uses gene interaction structure as prior biological information~\citep{txpert}, while STATE learns a transformer-based transition between cellular states~\citep{state}. CellFlow and PerturbDiff model response distributions through conditional flow matching and functional diffusion, respectively~\citep{cellflow,perturbdiff}. We retrain these four methods for our prediction tasks. Occam conditions a frozen cell representation on perturbation features, combining network neighborhoods with GenePT gene-text and ESM2 protein-sequence embeddings~\citep{genept,esm2}. The transfer baseline reuses the same perturbation's measured effects from other training cell lines; sum of singles adds measured single-gene effects.

\section{Method}
\label{sec:method}
\subsection{Representation learning}
\begin{figure}[!tp]
    \centering
    \includegraphics[width=0.98\linewidth]{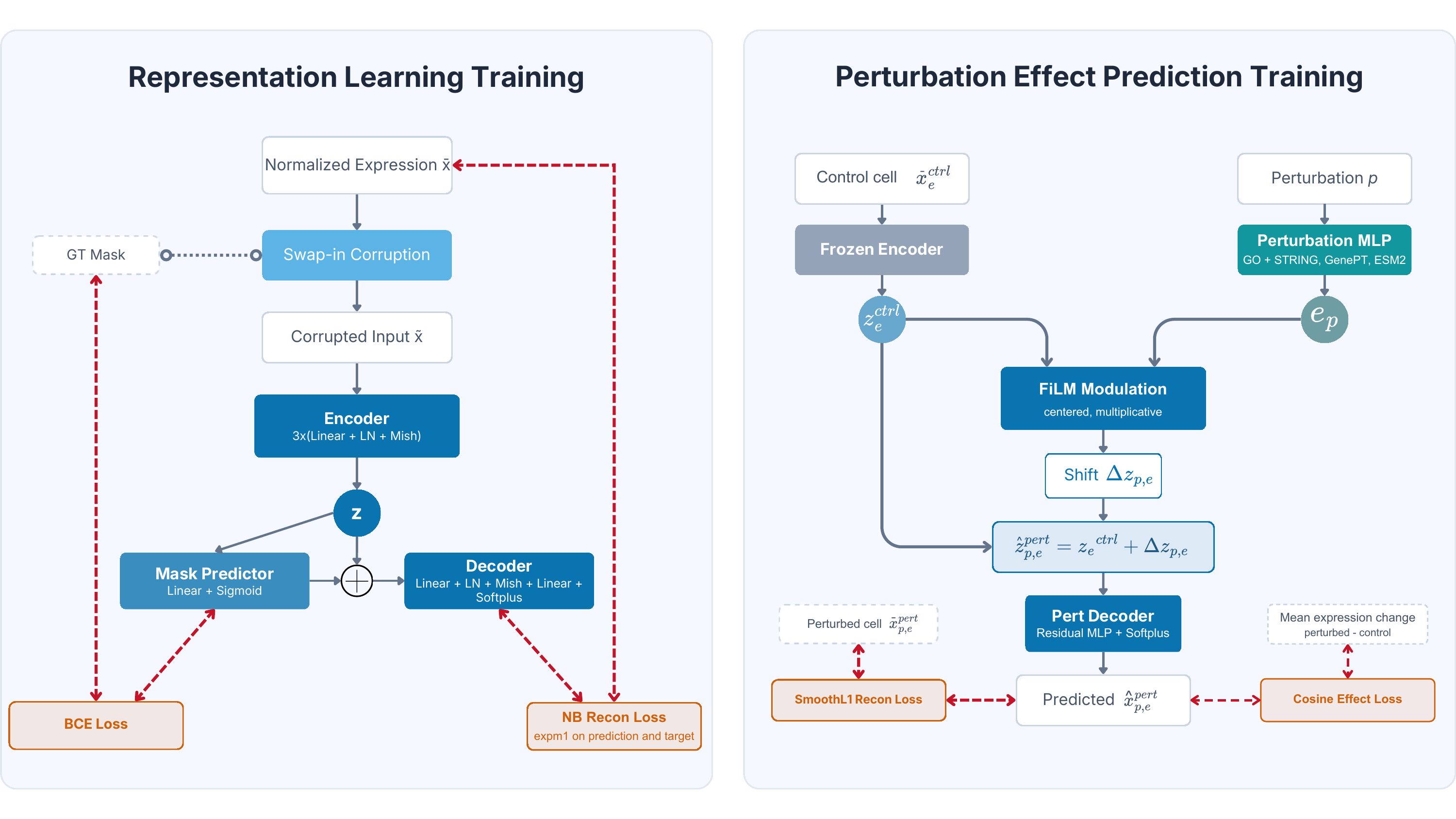}
    \caption{\textsc{Occam}'s training stages. Left, representation learning: reconstructing expression and detecting entries replaced from one donor cell train the encoder. The negative-binomial (NB) loss follows $\operatorname{expm1}$, which inverts the log transform. Right, perturbation-effect prediction: a frozen cell encoder, perturbation multilayer perceptron (MLP), modulation module, and trainable decoder predict perturbed expression from a control cell. Perturbation features describe gene networks (GO+STRING), gene text (GenePT), and protein sequences (ESM2). SmoothL1 supervises expression levels; a cosine loss aligns changes relative to controls with the observed mean effect.}
    \label{fig:architecture}
\vskip -0.15in 
\end{figure}

Occam uses multilayer perceptrons (MLPs) for representation learning and perturbation-effect prediction (Figure~\ref{fig:architecture}). Representation learning reconstructs expression from corrupted inputs. We normalize a nonnegative vector $\mathbf{x}\in\mathbb{R}_{\geq0}^{G}$ to a common total expression and apply $\log(1+\cdot)$, giving $\bar{\mathbf{x}}$. The encoder maps this $G=44{,}728$-gene profile to a $d=512$-dimensional representation.

\paragraph{Swap-in denoising.}
We replace a random subset of gene values with measurements of the same genes in another cell. For replacement probability $\rho$, a binary mask $\mathbf{m}_i\sim\operatorname{Bernoulli}(\rho)^{\otimes G}$ and a permutation $\pi$ of cells within the minibatch define $\tilde{\mathbf{x}}_i=\mathbf{m}_i\odot\bar{\mathbf{x}}_{\pi(i)}+(\mathbf{1}-\mathbf{m}_i)\odot\bar{\mathbf{x}}_i$, where $\odot$ denotes elementwise multiplication. All replaced entries come from the same donor cell. The model must therefore recover the recipient's profile from a mixture of two measured profiles. An MLP encoder produces $\mathbf{z}_i=f_\theta(\tilde{\mathbf{x}}_i)$. A sigmoid head estimates the replacement mask, $\hat{\mathbf{m}}_i=h_\psi(\mathbf{z}_i)$, and a decoder reconstructs expression from their concatenation, $[\mathbf{z}_i;\hat{\mathbf{m}}_i]$.

\paragraph{Reconstruction criterion.}
A softplus decoder output keeps predicted log-expression nonnegative. We invert the log transform with $\operatorname{expm1}$ for both predicted and target profiles, then compute the negative-binomial (NB) reconstruction loss in normalized count space. Encoder pretraining combines this loss with binary cross-entropy for mask prediction. Appendix~\ref{supp_training_details} gives the training schedule, and Appendix~\ref{app:softplus_count} derives the output parameterization.

\subsection{Perturbation-effect prediction}
\label{sec:virtual_cell}
The perturbation-effect predictor maps a control cell and a perturbation $p$ to perturbed log-expression in context $c$, here a cell line. We freeze the pretrained cell encoder and train a perturbation encoder, modulation module, and expression decoder on independently sampled control and perturbed cells from that line.

\paragraph{Perturbation features.}
We describe each target gene using neighborhoods in Gene Ontology (GO) and STRING networks, GenePT text embeddings, and ESM2 protein-sequence embeddings. Denoting these features by $\mathbf{a}_p$, $\mathbf{t}_p$, and $\mathbf{s}_p$, a shared MLP produces $\mathbf{e}_p=q_\eta([\mathbf{a}_p;\tilde{\mathbf{t}}_p;\tilde{\mathbf{s}}_p])$. Tildes denote normalization and scale matching of text and sequence features. Network neighbors remain binary adjacency features; the model performs no graph message passing. Appendix~\ref{supp_perturbation_embedding} specifies preprocessing and missing-feature handling.

\paragraph{Conditioning the cell state.}
The perturbation embedding modulates the frozen representation through multiplicative feature-wise linear modulation (FiLM)~\citep{film}. Let $\mathbf{z}=f_\theta(\bar{\mathbf{x}}^{\rm ctrl})$, let $\boldsymbol{\gamma}(\mathbf{e})$ be a sigmoid projection, and let $\bar{\mathbf{e}}$ be the mean perturbation embedding over the fixed target-gene vocabulary. We compute $\hat{\mathbf{z}}=\mathbf{z}+[\boldsymbol{\gamma}(\mathbf{e}_p)-\boldsymbol{\gamma}(\bar{\mathbf{e}})]\odot\mathbf{z}$. Subtracting the projection of the mean embedding centers the modulation on this reference; there is no additive FiLM shift. A trainable residual decoder with a softplus output predicts expression, $F_p(\bar{\mathbf{x}}^{\rm ctrl})=d_\omega(\hat{\mathbf{z}})$. For non-targeting controls, we bypass modulation and decode $\mathbf{z}$ directly.

\paragraph{Training objective.}
We supervise expression levels and perturbation-induced changes. Let $\hat{\mathbf{x}}=F_p(\bar{\mathbf{x}}^{\rm ctrl})$ be the prediction and $\bar{\mathbf{x}}^{\rm pert}$ an independently sampled perturbed target. Let $\boldsymbol{\mu}_{p,c}$ and $\boldsymbol{\mu}_{0,c}$ be training-population means of perturbed and control log-expression in context $c$. We minimize $\mathcal{L}_{\rm pert}=\operatorname{SmoothL1}(\hat{\mathbf{x}},\bar{\mathbf{x}}^{\rm pert})+\lambda_\Delta[1-\cos(\hat{\mathbf{x}}-\boldsymbol{\mu}_{0,c},\boldsymbol{\mu}_{p,c}-\boldsymbol{\mu}_{0,c})]$ over benchmark genes, where $\cos$ denotes cosine similarity and $\lambda_\Delta$ weights direction supervision. SmoothL1 compares the prediction with a sampled single-cell target; the cosine term aligns its change from controls with the measured mean effect. Both losses operate on decoded expression. Appendix~\ref{supp_training_details} gives the loss settings.

\paragraph{Predicting double perturbations.}
To predict simultaneous perturbation of genes $a$ and $b$, we compose the single-perturbation predictor on a log-expression input $\bar{\mathbf{x}}$: $F_{a,b}(\bar{\mathbf{x}})=\tfrac12[F_b(F_a(\bar{\mathbf{x}}))+F_a(F_b(\bar{\mathbf{x}}))]$. Each step decodes an expression profile and re-encodes it for the next perturbation. Both orders use the same weights; averaging makes the prediction independent of gene order. Parameter fitting uses single-gene perturbations. We use validation double perturbations to select checkpoints and a disjoint set of test double perturbations to assess generalization (Section~\ref{sec:training_setup}; Appendix~\ref{app:vc_cohorts}).

\section{Experiments}
\label{sec:experiments}
Representation evaluation measures information in embeddings of observed cells. Perturbation-effect prediction evaluates expression profiles generated for specified interventions. Prediction tests cover unobserved perturbation--cell-line combinations, cell lines held out from predictor training, and double perturbations that target two genes simultaneously.

\subsection{Training setup and data splits}
\label{sec:training_setup}
\paragraph{Representation pretraining.}
For the controlled representation comparison, encoders are pretrained on DiverseRNA~\citep{txfm}: approximately 1.4 million single-cell and bulk RNA-seq profiles from nine datasets. Occam reserves approximately 10\% of profiles for validation; studies and cell types can appear in both splits. Competitors use the same corpus with their own preprocessing and objectives. We also evaluate public checkpoints pretrained on other corpora. Appendices~\ref{supp_training_details} and~\ref{supp_end_of_training}--\ref{sec:scmae-training} detail training.

\paragraph{Perturbation-effect prediction.}
We train separate predictors on two CRISPR perturbation datasets. \textbf{RHJ} combines CRISPR interference screens, which knock down target-gene expression, in RPE1 (retinal epithelial), HepG2 (liver-derived), and Jurkat (T-cell-derived) cell lines~\citep{replogle,nadig}. We use 103 held-out HepG2 gene perturbations for validation and a disjoint set of 611 for testing. The model trains on other perturbations in HepG2 and on the held-out perturbation identities in Jurkat and/or RPE1. Thus, both the perturbation and the target cell line are observed during training, but their specific combination is not. We call this the \emph{cross-context few-shot} setting; no training cells receive a held-out perturbation in HepG2.

\textbf{Norman} profiles CRISPR activation in the leukemia-derived K562 cell line~\citep{norman2019}. Single perturbations activate one gene; \emph{double perturbations} activate two genes simultaneously in the same cell. Training uses 105 single-gene perturbation conditions (47,269 cells) and 9,702 non-targeting control cells. We reserve 20 double perturbations (5,276 cells) for validation and a disjoint set of 111 double perturbations (28,921 cells) for testing. Both constituent single-gene perturbations are present in training. Double-perturbation cells are used only for checkpoint selection and testing, with no gradient updates.

On each perturbation dataset, Occam keeps the pretrained cell encoder frozen and trains the perturbation modules and expression decoder with five random seeds. Each run retains the epoch with the highest validation \textbf{Pearson $\Delta$}, averaged over validation conditions. For Norman, we predict each validation and test double perturbation by applying the single-gene predictor in both orders (gene $a$ then $b$, and $b$ then $a$) and averaging the two resulting expression profiles (Section~\ref{sec:virtual_cell}). Appendix~\ref{supp_training_details} gives schedules and loss weights.

\subsection{Cell-representation evaluation}
The perturbation representation benchmark measures information retained in frozen cell embeddings using HepG2, Jurkat, and RPE1~\citep{bendidi2024benchmarkpca}. All three cell lines and their benchmark data are absent from DiverseRNA~\citep{txfm} and are therefore entirely unseen during encoder pretraining in the controlled comparison. We keep the model frozen and use linear probing for some metrics. Tasks assess perturbation identity, gene relationships, and expression information across batches. Each cell-line score averages six tasks; the overall score averages the three cell lines.

The scIB cell-integration benchmark~\citep{scib} tests whether embeddings mix cells from different technical batches while preserving cell types. We use pancreatic cells (Pancreas) and peripheral blood mononuclear cells (PBMC-68k)~\citep{pancreas1,pancreas2,pbmc68k}. Each dataset score weights biological conservation by 0.6 and batch integration by 0.4; the overall score averages the two datasets equally. Appendices~\ref{app:repr_datasets}--\ref{supp_scIBBenchmark} define metrics and preprocessing; Appendix~\ref{supp_gene_phenotype_eval} describes an additional gene-embedding phenotype benchmark.

\subsection{Perturbation-effect prediction tasks}
\label{sec:vc_evaluation}
\paragraph{Generalization across cell lines and assays.}
We evaluate three settings to test generalization to new perturbation--cell-line combinations and to new cell lines, and to assess prediction performance when the assay also changes. Seen and unseen refer to \emph{predictor training}.
\begin{enumerate}[leftmargin=*,nosep]
\item \textbf{Few-shot: RHJ/HepG2.} We evaluate 611 HepG2 gene perturbations. Both the perturbation and HepG2 are seen during training, but never together: the perturbation is observed in Jurkat and/or RPE1~\citep{replogle,nadig}.
\item \textbf{Unseen cell line, seen assay: K562.} We evaluate 528 perturbations in the Replogle K562 leukemia-cell knockdown screen~\citep{replogle}, which shares the assay family used for RHJ RPE1.
\item \textbf{Unseen cell line and assay: H1.} We evaluate 80 perturbations in the Virtual Cell Challenge H1 human embryonic stem-cell screen~\citep{vcc2025}.
\end{enumerate}
We apply the RHJ-trained predictors to K562 and H1 without fine-tuning. We also evaluate K562 on the same 80 gene perturbations as H1. This tests generalization across cell lines and assays without differences in which genes are perturbed (Appendix~\ref{app:vc_cohorts}).

\paragraph{Held-out double perturbations.}
We evaluate all 111 held-out double perturbations in the Norman activation dataset. For each, we compare the measured effect with the sum of the two measured single-gene effects. We group double perturbations by how strongly they change gene expression, then select responses closest to and furthest from this sum within each group. This yields 36 strongly additive and 36 weakly additive double perturbations, with small and large expression changes in both subsets. Overall results include all 111 double perturbations, including the 39 intermediate responses (Appendix~\ref{app:vc_cohorts}).

\newpage
\begin{table}[!tp]
    \centering
    \caption{\textbf{Cell-representation quality by dataset and overall.} Trained methods in the first two groups share the DiverseRNA corpus; public checkpoints use much larger corpora. Bold and underline mark the best and second-best mean, respectively, in each column across all groups.}
    \label{tab:representation}
    \label{tab:main_results}
    \fontsize{9}{10}\selectfont
    \setlength{\tabcolsep}{0.4pt}
    \begin{tabular*}{\linewidth}{@{\extracolsep{\fill}}l*{4}{c}@{\hspace{5pt}}*{3}{c}@{}}
    \toprule
     & \multicolumn{4}{c}{\textbf{Perturbation representation} $\uparrow$} & \multicolumn{3}{c}{\textbf{scIB} $\uparrow$} \\
    \cmidrule(lr){2-5}\cmidrule(lr){6-8}
    Model & RPE1 & HepG2 & Jurkat & \textbf{Overall} & Pancreas & PBMC-68k & \textbf{Overall} \\
    \midrule
    \multicolumn{8}{@{}l}{\makebox[84pt][l]{\textbf{Nonlinear encoders}}Trained on DiverseRNA (\textbf{1.4M profiles})} \\
    \rowcolor{occamrow}
    Occam-90M & \reprcell{\textbf{42.94}}{0.04} & \reprcell{\textbf{38.79}}{0.04} & \reprcell{\underline{36.22}}{0.02} & \reprcell{\textbf{39.31}}{0.02} & \reprcell{\textbf{53.10}}{0.12} & \reprcell{54.33}{0.16} & \reprcell{\textbf{53.72}}{0.10} \\
    TxFM-B & \reprcell{\underline{42.05}}{0.02} & \reprcell{\underline{37.50}}{0.02} & \reprcell{\textbf{36.81}}{0.01} & \reprcell{\underline{38.78}}{0.01} & \reprcell{47.29}{0.10} & \reprcell{53.14}{0.24} & \reprcell{50.22}{0.13} \\
    scVI-110M & \reprcell{35.55}{0.08} & \reprcell{33.29}{0.08} & \reprcell{30.62}{0.04} & \reprcell{33.16}{0.04} & \reprcell{48.72}{0.18} & \reprcell{\underline{55.02}}{0.24} & \reprcell{51.87}{0.15} \\
    STATE-SE-90M & \reprcell{35.86}{0.20} & \reprcell{32.14}{0.20} & \reprcell{30.78}{0.17} & \reprcell{32.93}{0.11} & \reprcell{47.03}{0.30} & \reprcell{48.38}{0.16} & \reprcell{47.71}{0.17} \\
    scMAE-90M & \reprcell{26.50}{0.38} & \reprcell{25.85}{0.37} & \reprcell{24.77}{0.34} & \reprcell{25.71}{0.21} & \reprcell{47.12}{0.24} & \reprcell{51.70}{0.18} & \reprcell{49.41}{0.15} \\
    scGPT-90M & \reprcell{24.25}{0.34} & \reprcell{24.31}{0.31} & \reprcell{23.47}{0.22} & \reprcell{24.01}{0.17} & \reprcell{37.39}{0.16} & \reprcell{53.66}{0.30} & \reprcell{45.53}{0.17} \\
    \midrule
    \multicolumn{8}{@{}l}{\makebox[84pt][l]{\textbf{Simple baselines}}Fitted on DiverseRNA (\textbf{1.4M profiles}); Random is untrained} \\
    Ridge-PCA & \reprcell{38.54}{0.16} & \reprcell{35.18}{0.16} & \reprcell{32.82}{0.08} & \reprcell{35.51}{0.08} & \reprcell{47.46}{0.24} & \reprcell{54.42}{0.10} & \reprcell{50.94}{0.13} \\
    PCA & \reprcell{38.34}{0.14} & \reprcell{35.05}{0.08} & \reprcell{32.59}{0.08} & \reprcell{35.33}{0.06} & \reprcell{47.48}{0.16} & \reprcell{54.56}{0.12} & \reprcell{51.02}{0.10} \\
    Linear AE (22M) & \reprcell{36.80}{0.16} & \reprcell{33.93}{0.16} & \reprcell{31.73}{0.08} & \reprcell{34.15}{0.08} & \reprcell{51.14}{0.18} & \reprcell{54.20}{0.24} & \reprcell{\underline{52.67}}{0.15} \\
    Random & \reprcell{19.37}{0.55} & \reprcell{19.28}{0.44} & \reprcell{19.15}{0.40} & \reprcell{19.26}{0.27} & \reprcell{43.59}{0.30} & \reprcell{47.71}{0.16} & \reprcell{45.65}{0.17} \\
    \midrule
    \multicolumn{8}{@{}l}{\makebox[84pt][l]{\textbf{Public checkpoints}}Pretrained on larger corpora (\textbf{33 to 167M cells})} \\
    STATE-SE-600M & \reprpoint{39.54} & \reprpoint{34.97} & \reprpoint{33.19} & \reprpoint{35.90} & \reprpoint{46.58} & \reprpoint{50.83} & \reprpoint{48.71} \\
    AIDO.Cell-100M & \reprpoint{34.56} & \reprpoint{32.21} & \reprpoint{30.90} & \reprpoint{32.56} & \reprpoint{44.95} & \reprpoint{53.23} & \reprpoint{49.09} \\
    TrFormer-sp & \reprpoint{31.50} & \reprpoint{28.57} & \reprpoint{27.11} & \reprpoint{29.06} & \reprpoint{\underline{52.22}} & \reprpoint{49.17} & \reprpoint{50.70} \\
    scGPT-43M & \reprpoint{25.20} & \reprpoint{25.01} & \reprpoint{24.06} & \reprpoint{24.75} & \reprpoint{41.89} & \reprpoint{\textbf{55.43}} & \reprpoint{48.66} \\
    \bottomrule
    \end{tabular*}
    \par\vspace{3pt}
    \begin{minipage}{\linewidth}\fontsize{8}{9}\selectfont
    Perturbation: mean of six tasks per dataset. scIB: cell-type conservation and batch mixing. Overall: mean of three perturbation datasets or two scIB datasets. M in model names: millions of parameters. TrFormer-sp: TranscriptFormer-Sapiens.
    \end{minipage}
    \end{table}
\subsection{Perturbation-effect prediction: comparators and scoring}
We retrain TxPert~\citep{txpert}, STATE~\citep{state}, CellFlow~\citep{cellflow}, and PerturbDiff~\citep{perturbdiff} on RHJ and Norman with five seeds, using the same evaluation conditions, genes, control draws, and scorer.

\paragraph{Simple baselines.}
\emph{Control cell} returns an observed control-cell profile; \emph{control mean} returns the average control-cell profile. \emph{Mean perturbation} adds the average training effect to a control profile, regardless of the target gene. \emph{Transfer} adds the same perturbation's effect measured in other training cell lines to a control profile from the evaluation cell line. For double perturbations, \emph{sum of singles} adds the two measured single-gene effects to a control profile. Appendices~\ref{supp_experimental_accuracy}--\ref{app:vc_competitors} detail baselines and training.

\paragraph{Split-half scoring for perturbation-effect prediction.}
Within each test perturbation, we divide measured cells into disjoint, equally sized halves A and B, stratified by batch; controls are split separately. \textbf{Half A is the common reference.} Comparing measured perturbed cells in B with those in A gives \emph{experimental reproducibility}. Models receive B controls and the perturbation identity, and generate as many perturbed profiles as there are measured cells in each half. These predictions are also compared with A, not with B's perturbed cells.

For both model evaluation and experimental reproducibility, an expression-change profile, or \emph{delta} ($\Delta$), is the mean perturbed profile minus the mean control profile. The common reference uses A's control mean. Measured B profiles and model predictions share the same measured B-control mean, keeping their scoring conditions matched. We use separate A and B controls because subtracting the same estimated control mean from both delta profiles introduces shared sampling noise that can inflate Pearson $\Delta$~\citep{nicol2026spurious,fukutani2026normalization}. Differentially expressed genes (DEGs) are likewise identified against A controls for the reference, and against B controls for both measured B cells and predictions. DEG scoring then compares each resulting gene list with A's reference list.

With more cells, smaller effects can become statistically significant, changing which genes are called DEGs. Matching cell counts and scored genes therefore prevents sample-size differences from being mistaken for differences in prediction quality. These evaluation splits are separate from training/validation/test assignment (Appendix~\ref{supp_experimental_accuracy}).

\begin{figure}[!tp]
\centering
\includegraphics[width=\linewidth]{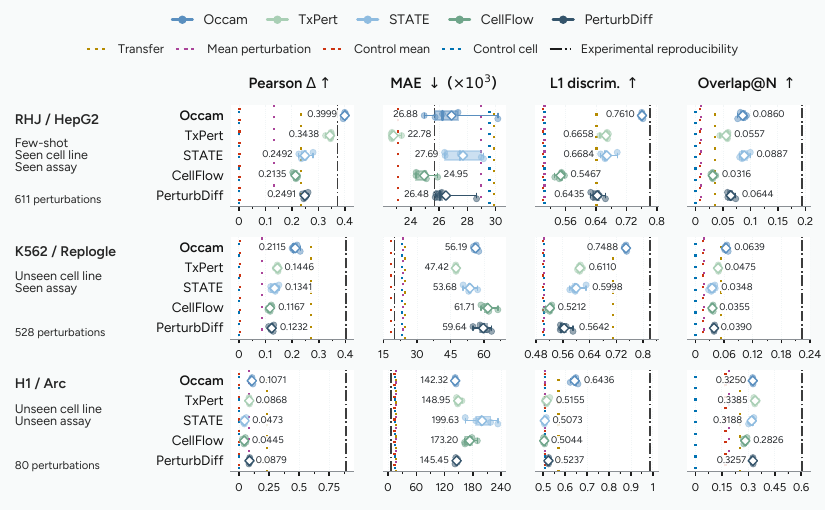}
\caption{\textbf{Perturbation-effect prediction across cell lines.} RHJ-trained models are evaluated on unseen perturbation--HepG2 combinations (few-shot), K562 (seen assay), and H1 (unseen assay), without fine-tuning. Transfer uses the same perturbation's measured effects from other training cell lines. Points show five training-seed means over five evaluation splits each. Boxes show quartiles and medians, whiskers show minima and maxima, and diamonds and labels show means. Dotted lines show simple-baseline means; dash-dot lines show experimental reproducibility (not a ceiling). Linear ranges vary by panel. MAE values are scaled by $10^3$. Means and SDs: Tables~\ref{tab:full:rhj_fewshot:all}--\ref{tab:full:k562_seen:all}.}
\label{fig:vc_context}
\end{figure}

\paragraph{Perturbation-effect prediction metrics.}
We use Cell-eval~\citep{celleval} for model evaluation and experimental reproducibility, without post-hoc calibration. Pearson $\Delta$ is Pearson's $r$ between a predicted or measured B delta and A's reference delta across genes, measuring agreement in their relative expression changes. Mean absolute error (MAE) compares mean log-normalized expression levels with A. L1 discrimination ranks A's measured responses by their L1 distance to the predicted or measured B delta and scores the rank of the correct perturbation (chance 0.5). Overlap@N measures recovery of A's DEGs. Each of five training seeds is evaluated on the same five distinct A/B splits, giving 25 seed--split combinations. We average scores over splits within each seed, then report the mean and sample standard deviation (SD) across the five seed means. For deterministic baselines, seeds change the sampled controls. Appendix~\ref{app:vc_metrics} gives metric definitions and uncertainty calculations.

\section{Results}
\label{sec:results}
\subsection{Cell-representation quality}
Occam leads both aggregate representation scores in Table~\ref{tab:representation}: 39.31 for perturbation information and 53.72 for scIB cell integration. Across all evaluated encoders and baselines, it leads on three of the five datasets: RPE1, HepG2, and Pancreas. TxFM leads on Jurkat; on PBMC-68k, scVI leads among models trained on DiverseRNA, while the public scGPT checkpoint has the highest score overall. Public checkpoints use different pretraining corpora containing approximately 33 to 167 million cells~\citep{scgpt,state}, compared with DiverseRNA's 1.4 million expression profiles, so the public-checkpoint comparison reflects differences in both data and architecture.

\begin{samepage}
Simple representations also perform well. Ridge-PCA reaches 35.51 on perturbation representation, ahead of scGPT, scMAE, STATE-SE, and scVI trained on DiverseRNA. Occam improves on Ridge-PCA by 3.80 points and exceeds TxFM's 38.78; the perturbation-score improvement over each comparator is significant ($p<0.01$, $n=5$). On scIB, the linear autoencoder is the closest comparator at 52.67, only 1.05 points behind Occam. Occam's margin over the strongest simple baseline is therefore larger on the perturbation benchmark. Tables~\ref{tab:supp_pert_repr_rpe1}--\ref{tab:supp_scib_PBMC-68k} give task-level scores.

\end{samepage}

\subsection{Perturbation-effect prediction across cell lines}
\paragraph{Consistent recovery of perturbation-specific responses.}
Across all three cell-line settings, Occam has the highest mean Pearson $\Delta$ among the five trained models and the highest mean L1 discrimination among all nine methods, including the four simple baselines (Figure~\ref{fig:vc_context}). Occam therefore better captures the relative expression changes across genes than the other trained models and better distinguishes perturbations than all other models and simple baselines. Across the four metrics, Occam leads in 8 of 12 comparisons among trained models and 5 of 12 including simple baselines. All model and baseline rankings exclude experimental reproducibility.

\paragraph{Unobserved combinations in a familiar cell line.}
On RHJ/HepG2, Occam leads the trained models and simple baselines on Pearson $\Delta$ and L1 discrimination. Its Pearson $\Delta$ is $0.3999\pm0.0035$, compared with $0.3438\pm0.0092$ for TxPert and 0.2332 for transferring the same perturbation's measured effects from other training cell lines. The discrimination gain is particularly clear: 0.761 versus 0.668 for STATE, the next-best method. Overlap@N is close to the best score (0.086 versus STATE's 0.089), while MAE is 0.0269 compared with TxPert's best value of 0.0228. Together, these results support stronger recovery of perturbation-specific response patterns with competitive DEG recovery and mean-expression accuracy. Its mean Pearson $\Delta$ also exceeds experimental reproducibility (0.3999 versus 0.3722).

\paragraph{Cell lines held out from predictor training.}
Without fine-tuning, Occam leads trained models on three metrics in both K562 and H1. On K562, where the assay is familiar, it leads on Pearson $\Delta$, L1 discrimination, and DEG recovery. On H1, where the assay is also unseen, it leads on Pearson $\Delta$, MAE, and discrimination, with DEG recovery close to TxPert's best score (0.325 versus 0.339). Occam's discrimination exceeds every simple baseline: 0.749 on K562 and 0.644 on H1, compared with 0.709 and 0.569 for transfer of the same perturbation's measured effects from other training cell lines. This transfer baseline has higher Pearson $\Delta$ in both settings, while control mean has lower MAE. All five deep learning models have high H1 MAE (0.1423--0.1996 versus 0.0137 for control mean), consistent with difficulty predicting expression levels under the combined cell-line and assay shift. Their positive Pearson $\Delta$ scores nevertheless indicate partial recovery of the relative expression changes across genes. Occam captures these response patterns best among the trained models and most clearly distinguishes perturbations, despite the shared difficulty predicting expression levels.

Matching target genes helps interpret the difference between K562 and H1. On the separate 80-perturbation K562 panel, Occam's Pearson $\Delta$ is $0.2033\pm0.0138$, versus $0.1071\pm0.0057$ for the same targets in H1. With the perturbation list held fixed, the lower H1 score is consistent with difficulty applying the predictor to an unseen assay, although differences in cell line, measured genes, and assay conditions prevent isolating the assay effect (Table~\ref{tab:full:k562_seen_h1set:all}; Appendix~\ref{app:vc_cohorts}).

\subsection{Perturbation-effect prediction for double perturbations}
\begin{figure}[!tp]
\centering
\includegraphics[width=\linewidth]{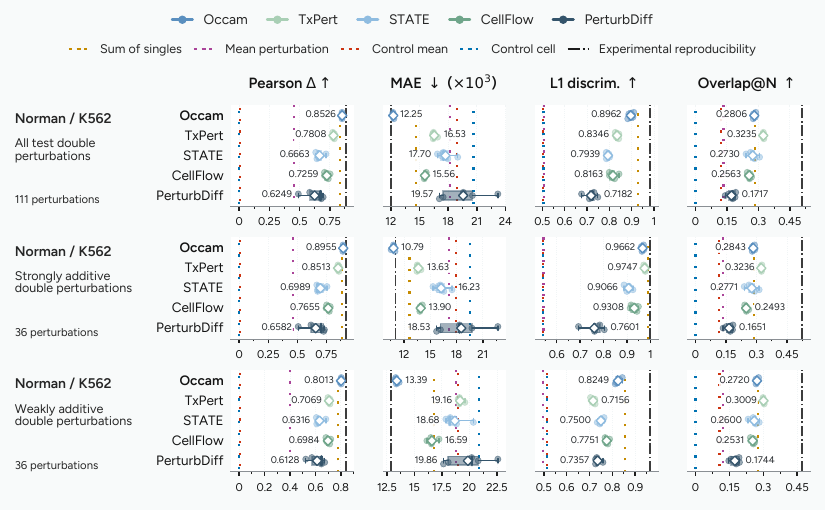}
\caption{\textbf{Perturbation-effect prediction for held-out double perturbations.} Norman's 111 test double perturbations include 36 strongly additive, 36 weakly additive, and 39 intermediate responses. Points show five training-seed means over five evaluation splits each. Boxes show quartiles and medians, whiskers show minima and maxima, and diamonds and labels show means. Dotted lines show simple-baseline means; dash-dot lines show experimental reproducibility (not a ceiling). Linear ranges vary by panel. MAE is scaled by $10^3$. Means and SDs: Tables~\ref{tab:full:norman_combo:additive}, \ref{tab:full:norman_combo:nonadditive}, and~\ref{tab:full:norman_combo:all}.}
\label{fig:vc_norman}
\end{figure}

\paragraph{Consistent gains in Pearson $\Delta$ and MAE.}
On Norman K562 double perturbations, Occam leads the trained models and simple baselines on Pearson $\Delta$ and MAE in the full test set and both additivity subsets (Figure~\ref{fig:vc_norman}). It has the best mean in 8 of 12 comparisons among trained models and 6 of 12 including simple baselines. The full set contains both subsets, so these are descriptive counts over overlapping evaluations. Across all 111 test doubles, Occam improves on sum of singles in Pearson $\Delta$ (0.8526 versus 0.8333) and MAE (0.01225 versus 0.01465), supporting the use of composed predictions beyond simply summing measured single-gene effects.

Occam's Pearson $\Delta$ advantage over sum of singles is larger for weakly additive doubles than for strongly additive doubles: 0.0264 versus 0.0133. MAE also improves in both subsets, indicating that the gains extend to responses less well described by adding single-gene effects. For strongly additive double perturbations, Occam achieves lower mean MAE than experimental reproducibility (0.01079 versus 0.01104). Each subset contains 36 double perturbations selected by departure from summed single-gene effects within effect-size groups. Because this selection uses the additive residual, the full-test-set comparison remains important for assessing gains over sum of singles.

\begin{samepage}
\paragraph{Discrimination and DEG recovery for double perturbations.}
Occam also leads trained models on L1 discrimination for all doubles and weakly additive doubles, and is close to TxPert on strongly additive doubles (0.9662 versus 0.9747). Sum of singles has the highest discrimination in all three evaluations. Occam's DEG recovery is close to sum of singles across the subsets; on the full test set, Overlap@N is 0.281 versus 0.283, with TxPert highest at 0.323. Thus, Occam improves Pearson $\Delta$ and MAE over sum of singles while recovering a similar proportion of reference DEGs. Sum of singles retains higher L1 discrimination.

\end{samepage}

\section{Conclusion}
\label{sec:discussion}
Occam establishes simple MLPs as a strong foundation for cell representation learning and perturbation-effect prediction. More complex architectures should be assessed by the gains in generalization they provide across biological conditions. The central challenge is to learn how perturbation effects depend on cellular context, so that predictions remain useful beyond the experimental systems used for training.

Broader coverage of cell contexts and perturbations should be a priority in both training and evaluation. Measuring the same perturbations across diverse contexts would help distinguish shared responses from context-specific effects. Matching perturbations and cell contexts across assays would help distinguish biological effects from technical variation. Training on this diversity should be accompanied by tests on unseen contexts, target genes, and perturbation combinations, including across assays. The next step is to establish how reliably models generalize across this wider experimental space, and whether their predictions can guide the choice of new experiments.

\clearpage
\begingroup
\setlength{\bibsep}{3pt}
\interlinepenalty=10000
\bibliographystyle{apalike}
\bibliography{references}
\endgroup

\clearpage
\appendix
\renewcommand{\thesection}{\Alph{section}}
\setcounter{section}{0}

\begin{center}
  {\LARGE\bfseries Appendix : Improved Single Cell Representation Learning and Perturbation Effect Prediction}\par
\end{center}
\vspace{1em}

\listofappendices

\let\oldsection\section
\let\oldsubsection\subsection

\let\origsection\section
\renewcommand{\section}[1]{%
  \origsection{#1}%
  \addcontentsline{apx}{section}{\protect\numberline{\thesection}#1}%
}
\let\origsubsection\subsection
\renewcommand{\subsection}[1]{%
  \origsubsection{#1}%
  \addcontentsline{apx}{subsection}{\protect\numberline{\thesubsection}#1}%
}

\section{Model limitations}
\label{sec:limitations}
Occam's encoder uses a fixed gene vocabulary. Applying it to another dataset requires matching gene identifiers and setting unmeasured genes to zero. The perturbation-effect predictor receives individual control cells, but the reported results mainly assess population-level effects. They do not establish accurate predictions of how a particular cell would respond or how responses vary between cells. GenePT, ESM2, and the gene networks supply prior information beyond the expression training data, and their coverage varies between genes.

Norman evaluates double perturbations whose constituent single-gene responses are observed during training. We do not evaluate unseen target genes, perturbations of more than two genes, or chemical perturbations. K562 and H1 differ in both cell line and measurement conditions, so their comparison cannot isolate the effect of either factor. Transferring measured effects from other training cell lines gives higher Pearson $\Delta$ than Occam in held-out cell lines, and no model leads on every metric. The five training seeds measure variation for the datasets and protocol studied here; the five evaluation splits per seed are not independent training runs. Validation in new prospective experiments remains future work.

\section{Output parameterization and count-space reconstruction}

\subsection{Softplus and count-space reconstruction}
\label{app:softplus_count}

Occam's representation decoder predicts log-normalized expression with a softplus output. We apply $\operatorname{expm1}$ to both predictions and targets before computing the reconstruction loss in normalized count space. The derivations below explain how this choice affects the gradients during encoder pretraining.

\paragraph{Setup.}
Let $r_{ij}\in\mathbb{R}$ denote the raw decoder output for gene $j$ in cell $i$, let
$t_{ij}=\bar{x}_{ij}=\log(1+y_{ij})$ denote the log-normalized target, and let
$y_{ij}=\operatorname{expm1}(t_{ij})$ denote the corresponding normalized count value. With a softplus output,
\begin{equation}
\hat{t}_{ij}
=
\operatorname{softplus}(r_{ij})
=
\log(1+\exp(r_{ij})).
\end{equation}
The count loss is evaluated after mapping this prediction back to normalized count space:
\begin{equation}
\hat{y}_{ij}
=
\operatorname{expm1}(\hat{t}_{ij})
=
\operatorname{expm1}(\log(1+\exp(r_{ij})))
=
\exp(r_{ij}).
\label{eq:softplus_lograte_identity}
\end{equation}
Applying softplus followed by $\operatorname{expm1}$ therefore gives $\hat y=\exp(r)$. The raw decoder output $r$ is the logarithm of the predicted normalized count mean; the softplus output is the predicted log-expression.

\paragraph{Gradients of the Poisson reconstruction loss.}
For a Poisson negative log-likelihood (NLL) reconstruction criterion in normalized count space, ignoring terms independent of the prediction,
\begin{equation}
\ell_{\mathrm{Pois}}(\hat{y},y)
=
\hat{y}
-
y\log \hat{y}.
\end{equation}
Using Eq.~\ref{eq:softplus_lograte_identity}, $\hat{y}=\exp(r)$, so
\begin{equation}
\ell_{\mathrm{Pois}}(r,y)
=
\exp(r)
-
yr,
\end{equation}
and therefore
\begin{equation}
\frac{\partial \ell_{\mathrm{Pois}}}{\partial r}
=
\exp(r)-y
=
\hat{y}-y,
\qquad
\frac{\partial^2 \ell_{\mathrm{Pois}}}{\partial r^2}
=
\exp(r)
=
\hat{y}.
\label{eq:poisson_lograte_grad}
\end{equation}
When a gene has zero measured expression, $y=0$, the gradient is $\hat y=\exp(r)$ and becomes small as the predicted count approaches zero. When the model underpredicts an expressed gene, $\hat y\ll y$, the gradient is approximately $-y$. Its magnitude therefore increases with the measured expression. These properties follow from the exponential count-mean parameterization in Eq.~\ref{eq:softplus_lograte_identity}.

\paragraph{Comparison with ReLU outputs.}
With a ReLU output,
\begin{equation}
\hat{t}=\max(0,r),
\qquad
\hat{y}=\operatorname{expm1}(\max(0,r)).
\end{equation}
For $r>0$, this gives $\hat{y}=\exp(r)-1$, whose optimum for target $y$ is
\begin{equation}
r^\star_{\mathrm{ReLU}}=\log(1+y).
\end{equation}
By contrast, under the softplus-count parameterization,
\begin{equation}
r^\star_{\mathrm{softplus}}=\log y,\qquad y>0.
\end{equation}

The difference is most pronounced for low positive counts. For example,
$y=0.1$ maps to $r^\star_{\mathrm{softplus}}\approx -2.30$ but
$r^\star_{\mathrm{ReLU}}\approx 0.095$. Thus ReLU followed by $\operatorname{expm1}$ compresses low positive normalized counts into a narrow region near zero, whereas softplus represents them on the natural log-rate scale.

\paragraph{Gradients with ReLU outputs.}
For $r<0$, ReLU has zero derivative. With a numerically stabilized Poisson loss, the gradient is therefore zero even when the measured count is positive. Without numerical stabilization, a zero predicted mean gives infinite loss for a positive target. For $r>0$, where $\hat y=\exp(r)-1$, the gradient
\begin{equation}
\frac{\partial \ell_{\mathrm{Pois}}}{\partial r}
=
\frac{\hat{y}+1}{\hat{y}}\,(\hat{y}-y)
\end{equation}
equals the softplus-count gradient $\hat y-y$ multiplied by $(\hat y+1)/\hat y$. This factor grows without bound as $\hat y$ approaches zero from above, which can produce very large gradients for small predicted counts.

\paragraph{Relation to NB and ZINB losses.}
The exact gradient in Eq.~\ref{eq:poisson_lograte_grad} applies to the Poisson objective. For a negative-binomial (NB) loss parameterized by mean $\mu$ and dispersion $\theta$, the negative log-likelihood has derivative with respect to the log-mean
\begin{equation}
\frac{\partial \ell_{\mathrm{NB}}}{\partial \log \mu}
=
\frac{\theta(\mu-y)}{\theta+\mu}.
\end{equation}
When $\mu=\exp(r)$, the raw output is again the log-mean. The NB gradient has the same sign as $\mu-y$, but its magnitude also depends on dispersion. It becomes small for zero targets when the predicted mean approaches zero. A zero-inflated negative-binomial (ZINB) loss includes an additional probability of zero expression, which also affects the gradient. NB and ZINB therefore use the same log-mean parameterization, but do not have the Poisson gradient above.

\paragraph{Why regression losses behave differently.}
Regression losses compare log-expression directly, without applying $\operatorname{expm1}$. For example, mean squared error (MSE) gives
\begin{equation}
\ell_{\mathrm{MSE}}
=
(\operatorname{softplus}(r)-t)^2,
\end{equation}
the gradient
\begin{equation}
\frac{\partial \ell_{\mathrm{MSE}}}{\partial r}
=
2\,(\operatorname{softplus}(r)-t)\,\sigma(r)
\end{equation}
approaches zero as $r\to-\infty$ because the sigmoid factor $\sigma(r)$ approaches zero. With a ReLU output, the corresponding gradient is zero for $r<0$. Neither activation produces the count-proportional gradient derived above when the loss compares log-expression directly.

\section{Implementation details}
\label{supp_model_details}

\subsection{Perturbation features and cell-state modulation}
\label{supp_perturbation_embedding}
\label{supp_model_architecture}
The perturbation-effect predictor maps a control cell and a target gene to perturbed log-expression. The pretrained cell encoder is frozen and has layer widths $44{,}728\to2048\to1024\to512$. The perturbation encoder represents 18,216 target genes. For each gene, we retain its top 20 neighbors in the Gene Ontology (GO) and STRING networks, make the connections symmetric, and encode each connection as a binary value. We add no self-connections and do not normalize by the number of neighbors. We concatenate these network indicators with a 3,072-dimensional GenePT text embedding and a 1,280-dimensional ESM2 protein-sequence embedding. The source embeddings remain fixed; the perturbation encoder and the vectors used for missing embeddings are trainable.

\paragraph{Source preprocessing.}
We align GenePT and ESM2 embeddings to the target-gene vocabulary and standardize each embedding dimension over genes with an available embedding. We then scale each source so that its average squared vector norm matches that of the network indicators. Let $E_A$ be the mean squared norm of nonzero network vectors and $E_k$ the mean squared norm of available, standardized embeddings from source $k$. We multiply those embeddings by $s_k=\sqrt{E_A/E_k}$ when both quantities are positive; otherwise, $s_k=1$. This keeps the sources on comparable scales while preserving differences in the number of network neighbors. Genes without an embedding receive a shared trainable vector for that source, initialized to zero. Their network indicators remain available.

An MLP maps the concatenated vector through Linear--ReLU--Dropout--Linear layers, with hidden width 128, output width 128, and dropout 0.2. It uses the network indicators directly, with no graph message passing. Let $\bar{\mathbf e}$ be the mean perturbation embedding over all target genes in the fixed vocabulary. A projection $g$ with widths $128\to256\to512$ and an identity hidden activation modulates the cell representation as
\begin{equation}
 \hat{\mathbf z}=\mathbf z+\left[\sigma(g(\mathbf e_p))-\sigma(g(\bar{\mathbf e}))\right]\odot\mathbf z .
 \label{eq:vc_modulation}
\end{equation}
Here, $\sigma$ is the sigmoid function. This multiplicative feature-wise linear modulation (FiLM) has scale one and no additive shift. Subtracting the projection of the mean perturbation embedding centers the modulation on that reference. Non-targeting controls bypass modulation. The expression decoder has hidden width 2048, four residual blocks with squeeze-and-excitation (SE) channel gating, SE reduction factor 16, dropout 0.3, and a softplus output. Reconstruction pretraining initializes its weights, which are then updated during perturbation-effect prediction training.

\subsection{Training stages and objective}
\label{supp_training_details}
\paragraph{Corpus and validation split.}
DiverseRNA~\citep{txfm} contains 1,434,299 expression profiles from nine public datasets, including seven single-cell datasets and bulk RNA-seq from GTEx and TCGA. Table~1 and Section~3 of \citet{txfm} describe the sources and curation. Occam uses a 44,728-gene input vocabulary. HepG2, Jurkat, and RPE1 and their representation-benchmark data are absent from DiverseRNA and are therefore unseen during encoder pretraining. The encoder remains frozen throughout representation evaluation.

For encoder pretraining, we divide the ordered corpus into groups of at most 8,192 profiles, shuffle the groups with seed 42, and reserve $\lfloor0.1K\rfloor$ of the $K$ groups for validation. This assigns approximately 10\% of profiles to validation. Training and validation contain different profiles, but can include the same studies and cell types. Validation measures reconstruction, separately from downstream perturbation evaluation. Representation baselines use their own preprocessing and training splits (Appendices~\ref{supp_end_of_training}--\ref{sec:scmae-training}).

\paragraph{Pretraining and reconstruction-decoder initialization.}
We normalize each DiverseRNA profile to a total count of 2,000 and apply $\log(1+\cdot)$. Encoder pretraining combines reconstruction in normalized count space with binary cross-entropy (BCE) for mask prediction. The encoder receives 100 epochs of negative-binomial (NB) pretraining with swap probability 0.1, learning rate $4\times10^{-4}$, batch size 1024, and mask-loss weight increasing from 0.7 to 0.9. This stage uses Adam and a cosine learning-rate schedule. It is followed by 100 epochs of Poisson reconstruction training at learning rate $6\times10^{-4}$, with gradually less input corruption and a mask-loss weight decreasing from 0.7 to zero. To initialize the perturbation-effect decoder, we keep the encoder frozen and train a separate expression decoder with NB reconstruction, learning rate $5\times10^{-5}$, batch size 1024, and a 200-epoch schedule. We use its epoch-124 checkpoint.

\paragraph{Perturbation-effect prediction targets.}
Each training example uses one control cell and one independently sampled perturbed cell from the same cell line. The control cell is the input; the perturbed cell is the reconstruction target. We also compute mean log-expression profiles from the training cells: $\boldsymbol\mu_{p,c}$ for perturbation $p$ in cell line $c$, and $\boldsymbol\mu_{0,c}$ for controls in that line. If $M$ selects the genes scored in the benchmark and $\hat{\mathbf x}_i$ is the predicted profile, the perturbation-effect prediction objective is
\begin{align}
 \mathcal L_{\rm pert}&=\frac{1}{B}\sum_i\left[\mathcal L_{\rm rec}^{(i)}+\lambda_\Delta\mathcal L_\Delta^{(i)}\right],\label{eq:pert_loss}\\
 \mathcal L_{\rm rec}^{(i)}&=\operatorname{SmoothL1}_{\beta=2}\!\left(M\odot\hat{\mathbf x}_i,M\odot\bar{\mathbf x}^{\rm pert}_i\right),\\
 \mathcal L_\Delta^{(i)}&=1-\cos\!\left(M\odot(\hat{\mathbf x}_i-\boldsymbol\mu_{0,c}),M\odot(\boldsymbol\mu_{p,c}-\boldsymbol\mu_{0,c})\right).
\end{align}
We average reconstruction loss over the benchmark genes. The cosine term aligns the predicted delta with the measured mean delta from training cells. Both losses operate on expression profiles, and no other loss terms are used. Training includes non-targeting controls and does not require paired measurements of a cell before and after perturbation. AdamW updates the expression decoder, perturbation encoder, and FiLM projection while the cell encoder remains frozen.

\paragraph{Training settings and checkpoint selection.}
RHJ training uses 180 epochs, batch size 2048, and direction-loss weight $\lambda_\Delta=0.5$. Norman training uses 150 epochs, batch size 32, and $\lambda_\Delta=0.0255$. Both use learning rate $10^{-3}$ for the trainable modules, five warmup epochs, gradient clipping at norm one, and training seeds 42--46. The cosine schedule reaches a minimum learning rate of $7\times10^{-4}$ for RHJ and $3\times10^{-4}$ for Norman. The Norman direction-loss weight approximately balances the average contributions of the two losses; the weights are specific to each training dataset. Each run retains the checkpoint with the highest validation Pearson $\Delta$, averaged over validation perturbations. We use the same RHJ checkpoints for HepG2, K562, and H1 evaluation, without fine-tuning. Norman is trained separately.

\paragraph{Predicting double perturbations.}
A double perturbation targets genes $a$ and $b$ simultaneously. Each single-perturbation predictor $F_p$ includes encoding, modulation, and decoding. We apply the two single-perturbation predictors in both orders and average the resulting expression profiles:
\begin{equation}
 F_{a,b}(\mathbf x)=\tfrac12\left[F_b(F_a(\mathbf x))+F_a(F_b(\mathbf x))\right].
 \label{eq:pair_compose}
\end{equation}
After the first prediction, the decoder output is passed back to the encoder as log-expression, without a second count normalization. Each order therefore uses two encoder and decoder passes. Both orders share the same weights. Averaging their final expression profiles makes the double-perturbation prediction independent of gene order.

\subsection{Cell-line-specific gene embeddings}
\label{supp_Occam_gene_embeddings}

The gene-representation benchmark~\citep{genbiobenchmark} trains phenotype probes on gene embeddings. Transformer models provide gene embeddings through their gene tokens. Occam is an MLP autoencoder, so it has no gene tokens. We instead combine its decoder weights with expression summaries from non-targeting control cells to construct one embedding per gene in each cell line. These embeddings are used only for the gene-representation benchmark, separately from the perturbation-effect predictor.

\paragraph{Aligned gene set and static decoder representation.}
Let $\mathbf W\in\mathbb R^{G\times h}$ contain the weights of the decoder's final linear layer. Each row $\mathbf w_g$ corresponds to gene $g$, and $h$ is the decoder hidden width. This benchmark uses its own decoder-weight embeddings; the perturbation-effect decoder described above has width 2048. For each cell line, we retain genes present in both the decoder vocabulary and the measured control cells:
\[
    \mathcal{G}_{\mathrm{obs}}
    =
    \mathcal{G}_{\mathrm{decoder}}
    \cap
    \mathcal{G}_{\mathrm{control}}.
\]
We construct embeddings only for genes $g\in\mathcal G_{\mathrm{obs}}$. The decoder-weight rows $\mathbf w_g$ provide static gene representations learned during reconstruction pretraining. The cell-line-specific information below comes only from control cells.

\paragraph{Control-derived expression summaries.}
We normalize non-targeting control profiles to a common total count and apply $\log(1+\cdot)$. Let $x_{ig}$ be the raw count of gene $g$ in control cell $i$, and let $\bar x_{ig}$ be its log-normalized value. Write $\mathcal B$ for the set of control batches and $N$ for the number of control cells. For each retained gene, we first compute its mean expression within batch $b$:
\[
    \mu_{b,g}
    =
    \frac{1}{N_b}
    \sum_{i:\,b_i=b}
    \bar{x}_{ig}.
\]
Here, $N_b$ is the number of control cells in batch $b$. We compute mean expression across batches ($\mu_g$), between-batch variance ($v_g$), and the ratio of mean expression to its between-batch standard deviation ($s_g$):
\begin{equation}
\label{eq:occam_gene_context_stats}
    \mu_g
    =
    \frac{1}{|\mathcal{B}|}
    \sum_{b\in\mathcal{B}}
    \mu_{b,g},
    \qquad
    v_g
    =
    \operatorname{Var}_{b\in\mathcal{B}}(\mu_{b,g}),
    \qquad
    s_g
    =
    \frac{\mu_g}{\sqrt{v_g}+\epsilon},
\end{equation}
We also compute the fraction of control cells with zero measured expression ($q_g$) and a variance-to-mean expression ratio ($d_g$). A small positive constant $\epsilon$ prevents division by zero:
\begin{equation}
\label{eq:occam_gene_dropout_dispersion}
    q_g
    =
    \frac{1}{N}
    \sum_{i=1}^{N}
    \mathbf{1}[x_{ig}=0],
    \qquad
    d_g
    =
    \frac{\operatorname{Var}_{i}(\bar{x}_{ig})}{\mu_g+\epsilon}.
\end{equation}
The basic control-summary vector is
\begin{equation}
\label{eq:occam_gene_basic_context}
    \mathbf{b}_g
    =
    \left[
    \mu_g,\ q_g,\ \log(1+d_g),\ \log(1+s_g)
    \right].
\end{equation}

\paragraph{Co-expression in control cells.}
We form a gene-by-cell matrix of log-normalized control expression and standardize each gene. Truncated singular value decomposition (SVD) with $32$ and $128$ components gives two sets of coordinates, $\mathbf a^{32}_g$ and $\mathbf a^{128}_g$, that summarize co-expression between genes. We concatenate these coordinates with the expression summaries above and standardize each dimension across genes:
\begin{equation}
\label{eq:occam_gene_context_target}
    \mathbf{t}_g
    =
    \operatorname{std}
    \left(
    \left[
    \mathbf{b}_g;\,
    \mathbf{a}^{32}_g;\,
    \mathbf{a}^{128}_g
    \right]
    \right),
    \qquad
    g\in\mathcal{G}_{\mathrm{obs}}.
\end{equation}
To smooth these expression summaries, we connect each gene to its 25 nearest neighbors by cosine distance. Let $\mathbf A_{\mathrm{expr}}$ be the row-normalized adjacency matrix of this graph, and let $\mathbf T$ contain one row $\mathbf t_g$ per gene. Three rounds of averaging over neighbors give
\begin{equation}
\label{eq:occam_gene_expr_diffusion}
    \tilde{\mathbf{T}}
    =
    \mathbf{A}_{\mathrm{expr}}^{3}
    \mathbf{T}.
\end{equation}

\paragraph{Smoothing with decoder weights and constructing the final embedding.}
The matrix $\tilde{\mathbf T}$ contains one smoothed expression-summary vector per gene. We build a second graph that connects each gene to its 30 nearest neighbors by cosine distance between decoder-weight rows $\mathbf w_g$. Let $\mathbf A_W$ be its row-normalized adjacency matrix.

We initialize the second smoothing step with the expression summaries from the first graph,
\[
    \mathbf{C}^{(0)}=\tilde{\mathbf{T}},
\]
and average over this second graph for five steps, retaining a contribution from the initial summaries at each step:
\begin{equation}
\label{eq:occam_gene_weight_diffusion}
    \mathbf{C}^{(k+1)}
    =
    \tau\mathbf{C}^{(0)}
    +
    (1-\tau)\mathbf{A}_{W}\mathbf{C}^{(k)},
    \qquad
    k=0,\dots,4,
\end{equation}
with $\tau=0.15$. The decoder weights determine which genes are neighbors; only the control-derived summaries are averaged. The final summary $\mathbf c_g$ for gene $g$ is the corresponding row of $\mathbf C^{(5)}$.

We apply a quantile transform to the smoothed mean-expression component of $\mathbf c_g$ to obtain a gene-specific scaling factor. This factor rescales the standardized decoder-weight row:
\[
    \eta_g
    =
    Q(c_{g,\mu}),
    \qquad
    \tilde{\mathbf{w}}_g
    =
    \eta_g\,\operatorname{std}(\mathbf{w}_g),
\]
where $Q$ is the quantile transform and $c_{g,\mu}$ is the smoothed mean-expression component. We concatenate the scaled decoder weights and expression summaries, standardize their dimensions, and apply $\ell_2$ normalization to obtain the final gene embedding:
\begin{equation}
\label{eq:occam_gene_embedding}
    \mathbf{u}_g
    =
    \operatorname{norm}_2
    \left(
    \left[
    \operatorname{std}(\tilde{\mathbf{w}}_g);
    \operatorname{std}(\mathbf{c}_g)
    \right]
    \right),
    \qquad
    g\in\mathcal{G}_{\mathrm{obs}},
\end{equation}
where $\operatorname{std}$ standardizes each embedding dimension across retained genes and $\operatorname{norm}_2$ divides each gene's vector by its $\ell_2$ norm.

\paragraph{Benchmark usage.}
This procedure gives one embedding per retained gene in each evaluated cell line. It uses the trained decoder weights and that cell line's non-targeting control cells. Perturbed-cell expression supplies the phenotype-probe targets, but is never used to construct the gene embeddings.

\section{Training the representation comparators}
\label{supp_competing_model_training}

\subsection{Training duration and checkpoint selection}
\label{supp_end_of_training}

All models in the controlled representation comparison use DiverseRNA and the same downstream benchmarks. Each model retains its own preprocessing, training objective, and training duration. For scGPT and scMAE, training stops after at least ten consecutive epochs without improvement in validation loss, and we retain the checkpoint with the lowest validation loss. STATE-SE uses the fixed schedule and checkpoint selection described below. The comparison therefore controls the training corpus, but not the amount of training compute.

\subsection{STATE-SE-90M training}
\label{sec:state-se-90m-training}

We train STATE's cell-embedding model, STATE-SE-90M, from scratch on DiverseRNA~\citep{txfm}, which contains $1{,}434{,}299$ expression profiles from nine datasets. We use 16 NVIDIA H100 80GB GPUs across two nodes, with eight GPUs per node. Training uses distributed data parallelism in PyTorch Lightning, NCCL communication between GPUs, and bfloat16 mixed precision. We follow \citet{state} with the settings below.

\paragraph{Architecture.}
The model uses a bidirectional transformer encoder with $L=16$ layers, hidden dimension $d_{\mathrm{model}}=512$, feed-forward dimension $d_{\mathrm{ffn}}=3584$, and $8$ attention heads, giving a head dimension of $64$. The output embedding dimension is $d_{\mathrm{out}}=512$. Each transformer block uses post-normalization, with residual connections followed by LayerNorm, and GELU activations in the feed-forward sub-layer. We compute scaled dot-product attention using Flash Attention in PyTorch. Dropout is set to $0.1$. The model contains approximately $83.6$M trainable parameters, including $75.6$M parameters in the transformer encoder.

\paragraph{Input representation.}
Each cell is represented as a sequence of at most $L_{\mathrm{seq}}=2048$ gene tokens. Gene identities are initialized from frozen ESM2 protein embeddings with dimension $d_{\mathrm{ESM2}}=5120$, covering a vocabulary of $19{,}790$ genes. A learnable \texttt{[CLS]} token is prepended to each sequence. Gene embeddings are $\ell_2$-normalized, projected to $d_{\mathrm{model}}$ with a linear layer followed by LayerNorm and SiLU, and then scaled by $\sqrt{d_{\mathrm{model}}}$.

\paragraph{Expression-count encoding.}
Expression values are incorporated using a soft-binning count encoder. Counts transformed with $\log(1+\cdot)$ are passed through a two-layer MLP with dimensions $1 \to 512 \to 10$ and LeakyReLU activation. The resulting 10-dimensional output defines soft assignments over learned count-bin embeddings of shape $10 \times d_{\mathrm{model}}$. The weighted sum of these count-bin embeddings is added to the corresponding projected gene-token embedding.

\paragraph{Dataset-classification auxiliary task.}
We append a trainable dataset token to each cell sequence to represent variation between source datasets. The encoded dataset-token representation is projected to a 10-dimensional embedding and passed to a dataset classifier consisting of Linear--SiLU--LayerNorm--Dropout--Linear layers. The classifier predicts the source dataset among the 9 training datasets using cross-entropy loss. This auxiliary dataset-classification loss is added to the main reconstruction objective.

\paragraph{Training objective.}
For each cell, the prediction task is constructed from sampled gene subsets. We sample $P=512$ positively expressed genes and $N=512$ non-expressed genes; the full task contains $1{,}536$ genes after including an additional set of 512 context/query genes. The main loss combines two maximum mean discrepancy (MMD) terms with energy kernels. The gene MMD term compares predicted and measured expression across cells for each gene. The cell MMD term compares predicted and measured expression across target genes for each cell. The total loss is
\begin{equation}
    \mathcal{L}
    =
    \mathcal{L}_{\mathrm{gene\_mmd}}
    +
    \mathcal{L}_{\mathrm{cell\_mmd}}
    +
    \mathcal{L}_{\mathrm{dataset}},
\end{equation}
where $\mathcal{L}_{\mathrm{dataset}}$ is the dataset-classification cross-entropy loss.

\paragraph{Sequencing depth and input masking.}
We condition predictions on sequencing depth using Raw Depth Adjustment (RDA). For each cell, we add its mean nonzero expression value to the decoder input. During training, we also apply gene masking with masking ratio $0.2$: masked gene-token inputs are zeroed out but remain included as prediction targets.

\paragraph{Optimization.}
We optimize the model with AdamW using peak learning rate $\eta_{\max}=10^{-5}$ and weight decay $0.01$. The learning-rate schedule begins with a linear warmup from $0.01\eta_{\max}$ to $\eta_{\max}$, followed by cosine annealing to $0.3\eta_{\max}$. The warmup phase lasts approximately $11{,}200$ optimizer steps. Gradients are clipped to maximum norm $0.8$. We use gradient accumulation over $8$ microbatches before each optimizer step.

\paragraph{Batch size and training length.}
The per-GPU microbatch size is $32$. With 16 GPUs and gradient accumulation over 8 microbatches, the effective batch size is
\begin{equation}
    32 \times 16 \times 8 = 4096
\end{equation}
cells per optimizer step. Each epoch contains approximately $350$ optimizer steps. The model is trained for $500$ epochs, corresponding to approximately $175{,}000$ optimizer steps.

\paragraph{Validation and checkpointing.}
Validation is performed every $1{,}200$ optimizer steps on $50$ minibatches from a held-out validation split. Checkpoints are saved every $500$ optimizer steps, and the top 4 checkpoints according to training loss are retained.

\paragraph{Reproducibility and data loading.}
All sampling uses random seed $42$. Training uses 16 data-loading workers per GPU, while validation uses 4 workers. Data loaders use persistent workers, pinned memory, and prefetch factor 4. 

\subsection{scGPT-90M training}
\label{sec:scgpt-90m}

We train scGPT-90M on DiverseRNA~\citep{txfm} and evaluate it on the same representation benchmarks as Occam. We retain scGPT's input representation and training objectives~\citep{scgpt}, while increasing the transformer size to match Occam's parameter count approximately. The resulting model has $91.7$M trainable parameters.

\paragraph{Architecture.}
The scGPT-90M model follows the scGPT architecture~\citep{scgpt}, with scaling applied only to the transformer encoder. The embedding dimension is kept fixed at $d_{\mathrm{model}}=512$, with $8$ attention heads and head dimension $64$. Relative to the published scGPT configuration, we increase the feed-forward hidden dimension from $512$ to $2048$ and the number of transformer layers from $12$ to $21$. This gives the model a standard $4\times$ feed-forward expansion ratio, while preserving the original tokenization scheme, expression-value encoder, \texttt{[CLS]} cell embedding, and decoder heads. The transformer encoder accounts for approximately $72\%$ of the parameters, the gene embedding table for approximately $25\%$, and the remaining parameters are contained in the expression, masked-value, and classification decoder heads.

\paragraph{Input representation and tokenization.}
Each cell is represented as a variable-length sequence of gene tokens paired with expression values. The vocabulary contains $44{,}731$ entries: $44{,}728$ genes and three special tokens, \texttt{<pad>}, \texttt{<cls>}, and \texttt{<eoc>}. A learnable \texttt{[CLS]} token is prepended to each sequence, and its final transformer representation is used as the 512-dimensional cell embedding. Expression values are quantile-binned into 51 discrete levels and encoded using the scGPT continuous value encoder. Sequences are truncated or padded to a maximum length of $1{,}200$ gene tokens.

\paragraph{Training objectives.}
We train scGPT-90M with the standard scGPT self-supervised objectives. Masked Expression Prediction (MEP) reconstructs masked expression values from token-level representations. Masked Value Prediction for Cell Embedding (MVC) reconstructs masked values from the \texttt{[CLS]} cell embedding, encouraging the cell embedding to capture whole-cell state. Gene Expression Prediction for Cell Embedding (GEPC) further regularizes the \texttt{[CLS]} representation by predicting gene expression values from the cell embedding. Zero-expression genes are modeled with the explicit zero-probability head used in scGPT.

\paragraph{Masking strategy.}
During each epoch, the expression masking ratio is sampled uniformly from $\{0.25,0.50,0.75\}$ and applied consistently across all cells for that epoch. The sampled mask ratio is broadcast across GPUs so that all distributed workers use the same corruption level. The corruption level therefore varies between epochs.

\paragraph{Training data.}
The model is trained on DiverseRNA~\citep{txfm}, containing $1{,}434{,}299$ expression profiles from nine datasets and $44{,}728$ genes. We normalize each profile to a total count of $10^4$ and apply $\log(1+\cdot)$. We assign 90\% of profiles to training and 10\% to validation. Profiles are read from disk and converted to gene tokens during loading, so the full expression matrix need not fit in memory.

\paragraph{Optimization.}
We optimize the model with Adam using learning rate $\eta_0=10^{-4}$ and no weight decay. The learning-rate schedule consists of a linear warmup over the first 10 epochs, increasing from $10^{-3}\eta_0$ to $\eta_0$, followed by StepLR decay with multiplicative factor $\gamma=0.996$ per epoch. Gradients are clipped to maximum norm $1.0$. Training uses automatic mixed precision with \texttt{bfloat16}/\texttt{float32} operations.

\paragraph{Distributed training and batch size.}
Training is distributed across 8 GPUs on a single node using PyTorch DistributedDataParallel. Each GPU processes a per-device batch size of $32$, giving an effective batch size of
\begin{equation}
    32 \times 8 = 256
\end{equation}
cells per optimizer step. Each worker receives a separate data shard through a distributed sampler with per-epoch reshuffling. One GPU process handles evaluation, checkpoint saving, and logging.

\paragraph{Training duration and checkpoint selection.}
Training follows the end-of-training criterion described in Appendix~\ref{supp_end_of_training}. Validation is performed every 2 epochs on the held-out split using the same masked reconstruction objective. Checkpoints are saved every 5 epochs, and the checkpoint with the lowest validation loss is retained for downstream evaluation.

\subsection{PCA, linear autoencoder, and scVI training}
\label{subsec:baseline-training}

PCA, Ridge-PCA, the linear autoencoder, and scVI use the same DiverseRNA corpus as Occam~\citep{txfm}: $1{,}434{,}299$ expression profiles, $44{,}728$ genes, and nine datasets. Each method selects its input genes as described below. Unless otherwise stated, input profiles were normalized to a target sum of $10^4$ counts per cell and transformed with $\log(1+x)$. Neural models use validation data for checkpoint selection as described below; PCA and Ridge-PCA fit directly to the training matrix. At inference, missing genes are zero-filled after alignment to each model's input vocabulary.

\paragraph{PCA.}
We fit principal component analysis (PCA) with $k=256$ components on the full preprocessed training matrix using the randomized SVD solver in scikit-learn. Randomized SVD provides an efficient low-rank approximation for matrices of this scale, with computational complexity approximately $\mathcal{O}(ndk)$ for $n$ cells, $d$ genes, and $k$ retained components. Cell embeddings are obtained by projecting centered expression profiles onto the learned principal subspace:
\begin{equation}
    \mathbf{z}
    =
    (\mathbf{x}-\boldsymbol{\mu})\mathbf{V}^{\top},
\end{equation}
where $\boldsymbol{\mu}$ is the training-set mean and $\mathbf{V}\in\mathbb{R}^{k\times d}$ contains the principal component loadings.

\paragraph{Ridge-PCA.}
Ridge-PCA uses the same $k=256$ components as PCA, with a ridge penalty on the projection. For a centered expression vector, the embedding is
\begin{equation}
    \mathbf{z}
    =
    (\mathbf{x}-\boldsymbol{\mu})
    \mathbf{V}^{\top}
    \left(
    \mathbf{V}\mathbf{V}^{\top}
    +
    \alpha\mathbf{I}
    \right)^{-1},
\end{equation}
with ridge penalty $\alpha=1.0$. For orthonormal PCA rows, this expression reduces to a uniform factor $1/(1+\alpha)$ on the PCA coordinates. It preserves neighbor ordering and does not define a different principal subspace; any metric difference must depend on scale-sensitive downstream processing or fitting variation. As with PCA, the embedding dimension is $256$.

\paragraph{Linear autoencoder.}
The linear autoencoder consists of a linear encoder
$f_\theta:\mathbb{R}^{d}\to\mathbb{R}^{k}$
and a linear decoder
$g_\phi:\mathbb{R}^{k}\to\mathbb{R}^{d}$,
with $d=44{,}728$ genes and latent dimension $k=256$. Both layers include bias terms, giving approximately $22.9$M trainable parameters. The model is trained to minimize mean squared reconstruction error,
\begin{equation}
    \mathcal{L}_{\mathrm{LAE}}
    =
    \left\|
    g_\phi(f_\theta(\mathbf{x}))-\mathbf{x}
    \right\|_2^2.
\end{equation}
Optimization uses Adam with initial learning rate $10^{-3}$, cosine annealing, and batch size $512$. Training data are streamed from disk in chunks of $8{,}192$ cells with double-buffered prefetching. Both encoder and decoder are linear. With MSE reconstruction and appropriate optimization, the learned subspace is closely related to the PCA subspace. Cell embeddings are the encoder outputs,
\begin{equation}
    \mathbf{z}=f_\theta(\mathbf{x}).
\end{equation}

\paragraph{scVI.}
We train scVI~\citep{scvi} using \texttt{scvi-tools} v1.3.3 on all $44{,}728$ genes. The model uses a variational autoencoder with one hidden layer of $512$ units in both the encoder and decoder, a $512$-dimensional latent space, and no dropout. The generative model uses a zero-inflated negative-binomial (ZINB) likelihood with gene-level dispersion parameters, and the latent prior is a standard normal distribution. The 9 constituent DiverseRNA datasets are provided as batch covariates, allowing the decoder to condition generation on dataset identity.

The model contains approximately $115.5$M parameters, including the latent encoder, library-size encoder, decoder, and gene-level dispersion parameters. Training optimizes the evidence lower bound (ELBO) with Adam at batch size $2{,}048$ on a single NVIDIA H100 GPU. The Kullback--Leibler (KL) divergence weight is annealed from 0 to 1 during training following the \texttt{scvi-tools} training schedule. The final checkpoint is selected by validation ELBO. Cell embeddings are defined as the posterior mean of the latent distribution,
\begin{equation}
    \mathbf{z}
    =
    \mathbb{E}_{q_\phi(\mathbf{z}\mid\mathbf{x})}[\mathbf{z}].
\end{equation}

\subsection{scMAE training}
\label{sec:scmae-training}

We train scMAE~\citep{scmae}, an MLP-based masked autoencoder, on DiverseRNA~\citep{txfm}. Its encoder and decoder learn to reconstruct expression, while an auxiliary head predicts which genes were corrupted. We use the same downstream representation benchmarks as for the other models.

\paragraph{Architecture.}
The encoder is a three-layer MLP that maps the input expression vector to a 512-dimensional latent representation. 
An auxiliary mask-prediction head, implemented as a single linear layer, maps the cell representation to one mask prediction per input gene. The decoder receives the concatenation of the latent representation and the predicted mask logits, and reconstructs the original expression profile through a linear output projection. The full model contains approximately $91$M trainable parameters.

\paragraph{Input preprocessing.}
We use scMAE's preprocessing. We select the top $8{,}000$ highly variable genes, normalize each cell by library size using the median total count across cells as the target depth, apply $\log(1+\cdot)$, and standardize each gene to zero mean and unit variance. The model is therefore trained on scaled log-normalized expression values rather than raw counts.

\paragraph{Training objective.}
During training, we apply swap-in corruption with masking probability $p=0.4$. The scMAE objective combines expression reconstruction with mask prediction. The reconstruction term is a weighted mean squared error between the reconstructed and original expression profiles. Corrupted genes receive weight $w_{\mathrm{masked}}=0.75$, while uncorrupted genes receive weight $w_{\mathrm{unmasked}}=0.25$, encouraging the model to focus on recovering replaced entries. The mask-prediction branch is trained with binary cross-entropy. The total loss is
\begin{equation}
    \mathcal{L}_{\mathrm{scMAE}}
    =
    \mathcal{L}_{\mathrm{recon}}
    +
    \lambda_{\mathrm{mask}}
    \mathcal{L}_{\mathrm{mask}},
\end{equation}
with $\lambda_{\mathrm{mask}}=0.7$.

\paragraph{Optimization.}
We train scMAE with Adam using learning rate $10^{-5}$, weight decay $1.82\times10^{-4}$, cosine learning-rate scheduling, and gradient clipping at maximum norm $1.0$. The batch size is $256$. Checkpoint selection follows the end-of-training criterion described in Appendix~\ref{supp_end_of_training}: the checkpoint with the lowest validation loss is retained, and training is stopped once the validation loss fails to improve for the specified patience window.

\section{Perturbation-effect prediction: benchmarks and comparators}
\label{app:vc_protocol}
\label{supp_perturbation_models_sources}
\subsection{Evaluation datasets and held-out perturbations}
\label{app:vc_cohorts}
\paragraph{RHJ: cross-context few-shot prediction.}
RHJ combines CRISPR interference screens, which knock down target-gene expression, in RPE1 (retinal epithelial), HepG2 (liver-derived), and Jurkat (T-cell-derived) cell lines. We use the HepG2 and Jurkat screens of \citet{nadig} and the RPE1 screen of \citet{replogle}, with 6,983 shared genes. Training uses 1,657 HepG2 gene perturbations; 103 different HepG2 perturbations are reserved for validation and 611 for testing. Each test perturbation has at least ten cells, with at least five cells per evaluation half. Validation and test perturbations are observed in Jurkat and/or RPE1 during training, while HepG2 is observed under other perturbations. Both the perturbation and the cell line are therefore familiar, but their combination is unseen. We call this the cross-context few-shot setting.

\paragraph{Cell lines held out from predictor training.}
We apply the RHJ-trained predictors to K562 leukemia cells and H1 human embryonic stem cells without fine-tuning. No K562 or H1 observations are used to train the perturbation-effect predictor. K562 is an unseen cell line measured with a seen assay; H1 is an unseen cell line measured with an unseen assay. Here, seen and unseen refer to predictor training. DiverseRNA includes K562 CRISPR interference profiles used for encoder pretraining~\citep{txfm}. K562 evaluation uses 528 RHJ test perturbations and 6,640 measured genes shared with RHJ. H1 evaluation uses 80 perturbations and 6,535 genes.

\paragraph{The same target genes in K562 and H1.}
We also evaluate K562 on the same 80 perturbations as H1, so that differences between these results cannot be attributed to the choice of target genes. Of these 80 perturbations, 56 are HepG2 training perturbations, 21 belong to the 611-perturbation HepG2 test set, and two are validation perturbations. The remaining perturbation, CHMP3, is held out from HepG2 training but has only three HepG2 cells and is not evaluated there. All 80 perturbations are observed in at least one RHJ training cell line. This 80-perturbation evaluation set is separate from, and not contained within, the 528-perturbation K562 test set. Models that require a cell-line label receive the HepG2 label for K562 and H1. All models use the same perturbations and sampled controls within each evaluation set. Cell line and measured genes still differ between the K562 and H1 evaluations.

\paragraph{Norman: held-out double perturbations.}
The Norman dataset~\citep{norman2019} uses CRISPR activation in K562 cells. Single perturbations activate one gene; double perturbations activate two genes simultaneously in the same cell. We reconstruct the dataset from GEO accession GSE133344 and retain 91,168 cells after quality control. We select 8,000 genes from those detected in more than 8\% of cells, also including perturbation targets with matched gene identifiers. Training uses 47,269 cells across 105 single-gene perturbations and 9,702 non-targeting control cells. We reserve 20 double perturbations (5,276 cells) for validation and a disjoint set of 111 double perturbations (28,921 cells) for testing. The counts of 20 and 111 refer to distinct double perturbations; cell counts are given in parentheses. Each run retains the checkpoint with the highest mean validation Pearson $\Delta$ across the 20 validation double perturbations. Both validation and test predictions apply the single-perturbation predictor in both orders and average the expression profiles (Appendix~\ref{supp_training_details}). Double-perturbation cells never contribute gradient updates.

RHJ and Norman control cells can be used in both training and evaluation. The A/B split below separates cells within evaluation; it does not exclude all training controls from evaluation.

\paragraph{Strongly and weakly additive double perturbations.}
We compare each measured double-perturbation effect with the sum of the two measured single-gene effects. A delta profile, $\Delta$, is mean perturbed log-expression minus mean control log-expression. For double perturbation $p$ targeting genes $a$ and $b$, we measure the departure from additivity as $r_p=\|\Delta_p-(\Delta_a+\Delta_b)\|_2/\|\Delta_p\|_2$. Dividing by the measured effect size, $\|\Delta_p\|_2$, expresses the departure relative to the strength of the double-perturbation response. These calculations use all measured cells for each perturbation, before A/B splitting.

We first divide double perturbations into three groups by effect size. Within each group, we divide them into three further groups by $r_p$. Taking the lowest and highest groups gives 36 strongly additive and 36 weakly additive double perturbations, with small and large effects represented in both subsets. The remaining 39 double perturbations form the intermediate subset. Quantile boundaries use linear interpolation; values exactly on either boundary remain in the intermediate group. All models are evaluated on the same subsets. We also report a second grouping based only on $r_p$, without first grouping by effect size; each of those three subsets contains 37 double perturbations. Results for the full test set include all 111 double perturbations. The groups are used only for analysis, not as model inputs. Because weakly additive responses are selected for their departure from summed single-gene effects, this subset is inherently difficult for the additive baseline. The full test set is needed to assess improvement over that baseline overall.

\paragraph{Genes and normalization in held-out cell lines.}
All methods are scored on the same genes within each evaluation set. RHJ genes not measured in K562 or H1 receive zero at model input and are excluded from scoring. For K562 and H1, normalization uses genes shared with the HepG2 dataset: 7,598 genes in K562 and 8,910 in H1. Scoring then uses the smaller sets of shared benchmark genes given above. RHJ training instead normalizes over all genes in each source dataset. Cell line, available genes, sequencing depth, normalization, and assay can therefore all differ between training and evaluation. The K562/H1 comparison does not isolate their separate effects. Our evaluation sets use the assignments described here, rather than the original Virtual Cell Challenge training and test split.

\subsection{Split-half scoring for perturbation-effect prediction}
\label{supp_experimental_accuracy}
This protocol evaluates perturbation-effect prediction against experimental reproducibility using the same cell counts, genes, and reference populations. Cell-representation evaluation is separate: the encoder stays frozen, and linear probes are trained for some metrics (Appendices~\ref{app:repr_datasets}--\ref{supp_scIBBenchmark}). With more cells, smaller effects can become statistically significant, changing which genes are called differentially expressed genes (DEGs). Comparing many generated profiles with fewer experimental profiles could therefore change DEG scores simply because of sample size. We match the reference cells, the number of profiles per perturbation, the scored genes, and the scoring method. This removes the difference in sample size; it does not establish that model outputs reproduce biological variability or are independent experimental replicates.
\paragraph{Splitting measured cells into A and B.}
The training, validation, and test assignments determine which perturbations are evaluated. We then use five random seeds to split the measured cells of each test perturbation into two halves. For perturbation $p$ with $n_p$ cells, each half contains $m_p=\lfloor n_p/2\rfloor$ cells. We shuffle cells within each batch and assign them alternately to A and B, continuing the alternation across batches. If $n_p$ is odd, we randomly remove one cell from the larger half. Controls are split separately using the same procedure. A and B share no cells within a split, although different splits reuse cells. Splitting within batches reduces batch imbalance, but cannot guarantee identical batch proportions when few cells are available. Each perturbation has the same cell count across splits and methods.

\paragraph{Common reference for model evaluation and experimental reproducibility.}
Half A contains the measured perturbed cells and controls used as the common reference. Experimental reproducibility compares measured B perturbed cells with this reference. Models receive B controls sampled uniformly with replacement, with one control per predicted profile. Their predictions are also compared with A, not with measured B perturbed cells. Both measured B and each model provide exactly $m_p$ perturbed profiles for perturbation $p$, matching A's cell count and scored genes. Individual cells are not paired across halves. All methods use the same evaluation splits and random seeds for sampling controls.

For scoring, both measured B perturbed cells and each model's predictions are evaluated using the complete set of measured B controls. This set can differ from the sampled controls given to a model as inputs. Model evaluation and experimental reproducibility therefore share exactly the same B controls and the same A reference.

\paragraph{Delta profiles and differential expression.}
An expression-change profile, or delta ($\Delta$), is a mean perturbed profile minus its corresponding mean control profile. Write $\bar x_{A,p}$ and $\bar x_{A,0}$ for the mean profiles of perturbation $p$ and controls in A, with analogous notation for B. The common reference delta is $\Delta_{A,p}=\bar x_{A,p}-\bar x_{A,0}$. For experimental reproducibility, the measured B delta is $\Delta_{B,p}=\bar x_{B,p}-\bar x_{B,0}$. For a model, the predicted delta is $\widehat\Delta_{B,p}=\overline{\hat x}_{B,p}-\bar x_{B,0}$. Pearson $\Delta$ compares either $\Delta_{B,p}$ or $\widehat\Delta_{B,p}$ with $\Delta_{A,p}$ across genes. A's control mean is therefore part of the common reference in both comparisons; B's control mean is shared by measured B and model predictions.

We identify DEGs by comparing perturbed profiles with the corresponding controls. The reference compares A perturbed cells with A controls. Experimental reproducibility compares B perturbed cells with B controls; model evaluation compares predicted profiles with those same B controls. Both resulting DEG lists are scored against A's DEG list. Cell-eval 0.6.0~\citep{celleval}, using pdex 0.1.24 for differential expression, computes these comparisons in log-normalized expression without post-hoc calibration. A's DEG results are computed once and reused for every comparison within that split.

\paragraph{Why the two halves use separate controls.}
Experimental reproducibility measures agreement when both perturbed cells and controls are independently sampled into disjoint halves. Subtracting one estimated control mean from both delta profiles introduces the same sampling noise into both vectors. This shared noise can inflate Pearson $\Delta$ even when predictions contain little perturbation-specific information~\citep{nicol2026spurious,fukutani2026normalization}. We use separate A and B controls to avoid this source of spurious correlation, while reusing B controls between model evaluation and experimental reproducibility. Both comparisons use the same A reference; within each comparison, the two deltas use distinct control samples.

\paragraph{Interpretation of experimental reproducibility.}
Experimental reproducibility measures agreement between two samples of cells receiving the same perturbation, at the cell counts used for evaluation. Both measured means contain sampling noise, while a model's estimate of the population mean can have lower variance. Experimental reproducibility is therefore not a strict performance ceiling. Exceeding it on one metric does not establish that a model is more accurate than the experiment. Scores should be compared within each evaluation set: changing the perturbations or measured genes can also change the scores.

\begin{description}[leftmargin=1.5em,style=nextline]
\item[Control mean and control cell.] Control mean returns the mean B-control profile for every perturbation. Control cell returns a sampled B-control profile. Neither baseline adds a perturbation effect.
\item[Mean perturbation.] Add the mean training perturbation profile minus the B-control mean to each sampled B control. Training profiles are averaged with equal weight per perturbation. For K562 and H1, we first add the mean HepG2 training delta to the mean of all controls from the evaluation cell line. We then subtract the B-control mean and add the result to each sampled B control. The prediction therefore includes the mean training delta and a small difference between the full control mean and the B-control mean. It does not depend on which gene is perturbed.
\item[Transfer.] Add the same perturbation's measured effect from other training cell lines to each sampled B control from the evaluation cell line. The source effects are weighted by their cell counts. This baseline uses perturbation identity but does not learn how the effect changes between cell lines.
\item[Sum of singles.] Add the two observed single-gene effects to each sampled B control; this is the additive baseline for double perturbations.
\item[Experimental reproducibility.] Compare measured half-B perturbed cells with the common half-A reference, using B and A controls, respectively. This is a measured comparison, not a predictive model or a performance ceiling.
\end{description}
We train each model with five seeds and evaluate each on the same five A/B splits. For simple baselines, the seed changes the sampled controls. We average scores over splits within each seed, then report the mean and sample SD across seeds (Appendix~\ref{app:vc_metrics}). Deterministic baselines can have zero SD. The 25 evaluations are not independent training runs. SDs describe variation between seed means, not confidence intervals, variation between individual cells, or tests of statistical significance.

\subsection{Retrained competitors}
\label{app:vc_competitors}
We retrain TxPert~\citep{txpert}, STATE~\citep{state}, CellFlow~\citep{cellflow}, and PerturbDiff~\citep{perturbdiff} on RHJ and Norman with training seeds 42--46. All models use the same evaluation perturbations, genes, sampled controls, and scoring method. The results describe these adapted implementations under our protocol; they are not the models' original published benchmark results.

\paragraph{TxPert.}
TxPert uses GO and STRING network information and receives one control cell as input, as do the other predictors. We use its validation-selected checkpoint. For double perturbations, it adds the two perturbation embeddings before predicting expression.
\begin{samepage}
\paragraph{STATE.}
We use STATE's state-transition model for perturbation-effect prediction, separately from its cell-embedding model used in representation evaluation. Its input sequences group cells by perturbation and cell line. For Norman double perturbations, we apply the single-perturbation predictor in both orders and average the final expression profiles. H1 predictions are scored without post-hoc calibration.

\end{samepage}
\paragraph{CellFlow.}
CellFlow uses flow matching with a PCA transformation fitted for each training run. It combines perturbation inputs using its own aggregation module, with two perturbation tokens for Norman double perturbations. We select checkpoints on validation data and transform predicted PCA coordinates back to expression over the scored genes.
\paragraph{PerturbDiff.}
PerturbDiff uses functional diffusion, with GenePT embeddings in the RHJ comparison and a separate model trained on Norman. We evaluate its predictions using the same sampled controls and scored genes as the other methods.

\subsection{Metrics and uncertainty}
\label{app:vc_metrics}
A mean profile here averages log-normalized expression over cells receiving the same perturbation. Its delta subtracts the corresponding mean control profile, as defined in Appendix~\ref{supp_experimental_accuracy}. For model evaluation, Pearson $\Delta$ compares the predicted delta with A's reference delta across genes. For experimental reproducibility, it compares the measured B delta with A's reference delta. Pearson $\Delta$ measures agreement in the relative expression changes across genes, not calibrated effect magnitude. Mean absolute error (MAE) compares mean log-normalized expression levels with A; lower values are better. L1 discrimination ranks A's perturbations by L1 distance to the predicted or measured B delta and scores the rank of the correct perturbation. Chance performance is 0.5.

Overlap@N measures recovery of A's reference DEGs. Cell-eval identifies DEGs using Wilcoxon tests with a false-discovery-rate threshold of 0.05. For each perturbation, $N$ is the number of significant genes in A. Significant genes in the measured B or model comparison are ranked by absolute fold change; the top $N$ are compared with A's $N$ DEGs. The score is the number of shared genes divided by $N$. If fewer than $N$ genes are significant in the measured B or model comparison, only those available genes can contribute. If either list is empty, the scorer returns zero. The same rule applies to model evaluation and experimental reproducibility. Statistical significance depends on both cell count and variation between cells. Matching cell counts makes the comparison fairer, but Overlap@N alone does not establish that predictions reproduce the full distribution of measured responses. Each reported score averages over perturbations in the evaluation set, not over paired individual cells.

For metric values $m_{s,k}$ at seed $s$ and split $k$, define $\bar m_s=\frac15\sum_{k=0}^4m_{s,k}$ and $\bar m=\frac15\sum_{s=42}^{46}\bar m_s$. Tables report $\bar m\pm\sqrt{\frac14\sum_s(\bar m_s-\bar m)^2}$. 

\paragraph{Reading the result figures and tables.}
Figures~\ref{fig:vc_context} and~\ref{fig:vc_norman} show seed means as points, medians and quartiles as boxes, and minima and maxima as whiskers. Diamonds and labels show means. Dotted lines show simple-baseline means; dash-dot lines show experimental reproducibility. Model order is fixed in figures and tables. Figure MAE is multiplied by $10^3$; table MAE is unscaled. Results compare mean scores among trained models and with simple baselines, excluding experimental reproducibility. Rankings are not significance tests; metrics and evaluation sets are not combined.

\clearpage
\section{Complete perturbation-effect prediction results}
\label{app:vc_full_results}
These tables report all four perturbation-effect prediction metrics for 11 evaluation sets: few-shot prediction in RHJ/HepG2, prediction in K562 and H1 without fine-tuning, and Norman double perturbations, including the subsets defined in Appendix~\ref{app:vc_cohorts}. Each of the 110 model, baseline, or experimental-reproducibility rows summarizes five seeds and five evaluation splits per seed, for 2,750 evaluations in total. Means and sample SDs are computed across the five seed means, as described in Appendix~\ref{app:vc_metrics}.
\begin{table}[!tp]
\centering
\caption{Perturbation-effect prediction: RHJ/HepG2, cross-context few-shot prediction (611 perturbations). Values are mean $\pm$ SD across five seed means, each averaged over five evaluation splits (Appendix~\ref{app:vc_metrics}).}
\label{tab:full:rhj_fewshot:all}
\small
\setlength{\tabcolsep}{3pt}
\begin{tabular}{@{}lcccc@{}}
\toprule
Model & Pearson $\Delta\uparrow$ & MAE $\downarrow$ & Overlap@N $\uparrow$ & \shortstack{L1 discrimination\\$\uparrow$} \\
\midrule
\textbf{Occam} & $0.3999\pm0.0035$ & $0.02688\pm0.00203$ & $0.0860\pm0.0066$ & $0.7610\pm0.0018$ \\
TxPert & $0.3438\pm0.0092$ & $0.02278\pm0.00032$ & $0.0557\pm0.0064$ & $0.6658\pm0.0090$ \\
STATE & $0.2492\pm0.0205$ & $0.02769\pm0.00134$ & $0.0887\pm0.0080$ & $0.6684\pm0.0164$ \\
CellFlow & $0.2135\pm0.0084$ & $0.02495\pm0.00063$ & $0.0316\pm0.0034$ & $0.5467\pm0.0098$ \\
PerturbDiff & $0.2491\pm0.0080$ & $0.02648\pm0.00123$ & $0.0644\pm0.0063$ & $0.6435\pm0.0152$ \\
Transfer & $0.2332\pm0.0005$ & $0.02986\pm0.00001$ & $0.0353\pm0.0004$ & $0.6416\pm0.0010$ \\
Mean perturbation & $0.1336\pm0.0008$ & $0.02896\pm0.00002$ & $0.0105\pm0.0003$ & $0.5047\pm0.0007$ \\
Control mean & $0.0016\pm0.0000$ & $0.02309\pm0.00000$ & $0.0081\pm0.0000$ & $0.5014\pm0.0000$ \\
Control cell & $0.0000\pm0.0012$ & $0.02957\pm0.00002$ & $0.0000\pm0.0000$ & $0.5046\pm0.0006$ \\
\midrule
Experimental reproducibility & $0.3722\pm0.0000$ & $0.02567\pm0.00000$ & $0.1942\pm0.0000$ & $0.7821\pm0.0000$ \\
\bottomrule
\end{tabular}
\end{table}

\begin{table}[!tp]
\centering
\caption{Perturbation-effect prediction: H1, unseen cell line and unseen assay (80 perturbations). Values are mean $\pm$ SD across five seed means, each averaged over five evaluation splits (Appendix~\ref{app:vc_metrics}).}
\label{tab:full:h1_seen:all}
\small
\setlength{\tabcolsep}{3pt}
\begin{tabular}{@{}lcccc@{}}
\toprule
Model & Pearson $\Delta\uparrow$ & MAE $\downarrow$ & Overlap@N $\uparrow$ & \shortstack{L1 discrimination\\$\uparrow$} \\
\midrule
\textbf{Occam} & $0.1071\pm0.0057$ & $0.14232\pm0.00121$ & $0.3250\pm0.0012$ & $0.6436\pm0.0132$ \\
TxPert & $0.0868\pm0.0015$ & $0.14895\pm0.00448$ & $0.3385\pm0.0069$ & $0.5155\pm0.0094$ \\
STATE & $0.0473\pm0.0102$ & $0.19963\pm0.02696$ & $0.3188\pm0.0126$ & $0.5073\pm0.0044$ \\
CellFlow & $0.0445\pm0.0114$ & $0.17320\pm0.01123$ & $0.2826\pm0.0089$ & $0.5044\pm0.0032$ \\
PerturbDiff & $0.0879\pm0.0025$ & $0.14545\pm0.00247$ & $0.3257\pm0.0015$ & $0.5237\pm0.0011$ \\
Transfer & $0.2324\pm0.0012$ & $0.01749\pm0.00001$ & $0.2542\pm0.0003$ & $0.5695\pm0.0003$ \\
Mean perturbation & $0.0846\pm0.0013$ & $0.01621\pm0.00000$ & $0.1882\pm0.0004$ & $0.5097\pm0.0007$ \\
Control mean & $0.0037\pm0.0000$ & $0.01374\pm0.00000$ & $0.1701\pm0.0000$ & $0.5082\pm0.0000$ \\
Control cell & $0.0007\pm0.0035$ & $0.01468\pm0.00001$ & $0.0000\pm0.0000$ & $0.5091\pm0.0004$ \\
\midrule
Experimental reproducibility & $0.8881\pm0.0000$ & $0.00649\pm0.00000$ & $0.6054\pm0.0000$ & $0.9870\pm0.0000$ \\
\bottomrule
\end{tabular}
\end{table}

\clearpage
\begin{table}[!tp]
\centering
\caption{Perturbation-effect prediction: K562, unseen cell line and seen assay (528 perturbations). Values are mean $\pm$ SD across five seed means, each averaged over five evaluation splits (Appendix~\ref{app:vc_metrics}).}
\label{tab:full:k562_seen:all}
\small
\setlength{\tabcolsep}{3pt}
\begin{tabular}{@{}lcccc@{}}
\toprule
Model & Pearson $\Delta\uparrow$ & MAE $\downarrow$ & Overlap@N $\uparrow$ & \shortstack{L1 discrimination\\$\uparrow$} \\
\midrule
\textbf{Occam} & $0.2115\pm0.0125$ & $0.05619\pm0.00094$ & $0.0639\pm0.0031$ & $0.7488\pm0.0029$ \\
TxPert & $0.1446\pm0.0048$ & $0.04742\pm0.00025$ & $0.0475\pm0.0017$ & $0.6110\pm0.0055$ \\
STATE & $0.1341\pm0.0138$ & $0.05368\pm0.00232$ & $0.0348\pm0.0060$ & $0.5998\pm0.0183$ \\
CellFlow & $0.1167\pm0.0067$ & $0.06171\pm0.00285$ & $0.0355\pm0.0017$ & $0.5212\pm0.0079$ \\
PerturbDiff & $0.1232\pm0.0076$ & $0.05964\pm0.00319$ & $0.0390\pm0.0008$ & $0.5642\pm0.0164$ \\
Transfer & $0.2710\pm0.0003$ & $0.02454\pm0.00001$ & $0.0533\pm0.0004$ & $0.7090\pm0.0008$ \\
Mean perturbation & $0.0870\pm0.0007$ & $0.02379\pm0.00001$ & $0.0184\pm0.0001$ & $0.5045\pm0.0003$ \\
Control mean & $0.0040\pm0.0000$ & $0.01845\pm0.00000$ & $0.0153\pm0.0000$ & $0.5018\pm0.0000$ \\
Control cell & $0.0004\pm0.0007$ & $0.02332\pm0.00001$ & $0.0000\pm0.0000$ & $0.5043\pm0.0003$ \\
\midrule
Experimental reproducibility & $0.4011\pm0.0000$ & $0.01993\pm0.00000$ & $0.2236\pm0.0000$ & $0.8206\pm0.0000$ \\
\bottomrule
\end{tabular}
\end{table}

\begin{table}[!tp]
\centering
\caption{Perturbation-effect prediction: K562, the same 80 target genes as H1. Values are mean $\pm$ SD across five seed means, each averaged over five evaluation splits (Appendix~\ref{app:vc_metrics}).}
\label{tab:full:k562_seen_h1set:all}
\small
\setlength{\tabcolsep}{3pt}
\begin{tabular}{@{}lcccc@{}}
\toprule
Model & Pearson $\Delta\uparrow$ & MAE $\downarrow$ & Overlap@N $\uparrow$ & \shortstack{L1 discrimination\\$\uparrow$} \\
\midrule
\textbf{Occam} & $0.2033\pm0.0138$ & $0.05532\pm0.00082$ & $0.0715\pm0.0045$ & $0.6981\pm0.0048$ \\
TxPert & $0.1323\pm0.0067$ & $0.04672\pm0.00024$ & $0.0467\pm0.0021$ & $0.6071\pm0.0029$ \\
STATE & $0.1143\pm0.0131$ & $0.05221\pm0.00260$ & $0.0331\pm0.0069$ & $0.5466\pm0.0176$ \\
CellFlow & $0.0857\pm0.0118$ & $0.05916\pm0.00344$ & $0.0315\pm0.0021$ & $0.5211\pm0.0045$ \\
PerturbDiff & $0.1101\pm0.0100$ & $0.05841\pm0.00329$ & $0.0368\pm0.0015$ & $0.5413\pm0.0065$ \\
Transfer & $0.2819\pm0.0012$ & $0.02238\pm0.00003$ & $0.0602\pm0.0014$ & $0.6627\pm0.0018$ \\
Mean perturbation & $0.0765\pm0.0021$ & $0.02226\pm0.00003$ & $0.0181\pm0.0005$ & $0.5108\pm0.0019$ \\
Control mean & $0.0045\pm0.0000$ & $0.01684\pm0.00000$ & $0.0174\pm0.0000$ & $0.5078\pm0.0000$ \\
Control cell & $0.0004\pm0.0034$ & $0.02156\pm0.00002$ & $0.0000\pm0.0000$ & $0.5113\pm0.0019$ \\
\midrule
Experimental reproducibility & $0.4149\pm0.0000$ & $0.01861\pm0.00000$ & $0.2643\pm0.0000$ & $0.8147\pm0.0000$ \\
\bottomrule
\end{tabular}
\end{table}

\clearpage
\begin{table}[!tp]
\centering
\caption{Perturbation-effect prediction: Norman, held-out double perturbations in K562. 36 strongly additive double perturbations, grouped by effect size. Values are mean $\pm$ SD across five seed means, each averaged over five evaluation splits (Appendix~\ref{app:vc_metrics}).}
\label{tab:full:norman_combo:additive}
\small
\setlength{\tabcolsep}{3pt}
\begin{tabular}{@{}lcccc@{}}
\toprule
Model & Pearson $\Delta\uparrow$ & MAE $\downarrow$ & Overlap@N $\uparrow$ & \shortstack{L1 discrimination\\$\uparrow$} \\
\midrule
\textbf{Occam} & $0.8955\pm0.0032$ & $0.01079\pm0.00016$ & $0.2843\pm0.0065$ & $0.9662\pm0.0046$ \\
TxPert & $0.8513\pm0.0080$ & $0.01363\pm0.00019$ & $0.3236\pm0.0021$ & $0.9747\pm0.0031$ \\
STATE & $0.6989\pm0.0385$ & $0.01623\pm0.00083$ & $0.2771\pm0.0277$ & $0.9066\pm0.0143$ \\
CellFlow & $0.7655\pm0.0154$ & $0.01390\pm0.00012$ & $0.2493\pm0.0083$ & $0.9308\pm0.0153$ \\
PerturbDiff & $0.6582\pm0.0960$ & $0.01853\pm0.00285$ & $0.1651\pm0.0207$ & $0.7601\pm0.0415$ \\
Sum of singles & $0.8822\pm0.0003$ & $0.01262\pm0.00001$ & $0.2864\pm0.0012$ & $0.9890\pm0.0003$ \\
Mean perturbation & $0.4631\pm0.0032$ & $0.01713\pm0.00005$ & $0.1344\pm0.0009$ & $0.5438\pm0.0044$ \\
Control mean & $0.0033\pm0.0000$ & $0.01793\pm0.00000$ & $0.1171\pm0.0000$ & $0.5432\pm0.0000$ \\
Control cell & $-0.0023\pm0.0096$ & $0.01954\pm0.00005$ & $0.0000\pm0.0000$ & $0.5428\pm0.0019$ \\
\midrule
Experimental reproducibility & $0.9165\pm0.0000$ & $0.01104\pm0.00000$ & $0.5247\pm0.0000$ & $0.9978\pm0.0000$ \\
\bottomrule
\end{tabular}
\end{table}

\begin{table}[!tp]
\centering
\caption{Perturbation-effect prediction: Norman, held-out double perturbations in K562. 36 weakly additive double perturbations, grouped by effect size. Values are mean $\pm$ SD across five seed means, each averaged over five evaluation splits (Appendix~\ref{app:vc_metrics}).}
\label{tab:full:norman_combo:nonadditive}
\small
\setlength{\tabcolsep}{3pt}
\begin{tabular}{@{}lcccc@{}}
\toprule
Model & Pearson $\Delta\uparrow$ & MAE $\downarrow$ & Overlap@N $\uparrow$ & \shortstack{L1 discrimination\\$\uparrow$} \\
\midrule
\textbf{Occam} & $0.8013\pm0.0044$ & $0.01339\pm0.00014$ & $0.2720\pm0.0048$ & $0.8249\pm0.0123$ \\
TxPert & $0.7069\pm0.0069$ & $0.01916\pm0.00024$ & $0.3009\pm0.0020$ & $0.7156\pm0.0083$ \\
STATE & $0.6316\pm0.0303$ & $0.01868\pm0.00101$ & $0.2600\pm0.0180$ & $0.7500\pm0.0114$ \\
CellFlow & $0.6984\pm0.0168$ & $0.01659\pm0.00040$ & $0.2531\pm0.0090$ & $0.7751\pm0.0086$ \\
PerturbDiff & $0.6128\pm0.0627$ & $0.01986\pm0.00203$ & $0.1744\pm0.0213$ & $0.7357\pm0.0133$ \\
Sum of singles & $0.7749\pm0.0012$ & $0.01679\pm0.00004$ & $0.2751\pm0.0011$ & $0.8567\pm0.0029$ \\
Mean perturbation & $0.4023\pm0.0028$ & $0.01879\pm0.00004$ & $0.1225\pm0.0009$ & $0.4980\pm0.0021$ \\
Control mean & $0.0055\pm0.0000$ & $0.01894\pm0.00000$ & $0.1065\pm0.0000$ & $0.5135\pm0.0000$ \\
Control cell & $-0.0013\pm0.0109$ & $0.02087\pm0.00005$ & $0.0000\pm0.0000$ & $0.5159\pm0.0012$ \\
\midrule
Experimental reproducibility & $0.8377\pm0.0000$ & $0.01289\pm0.00000$ & $0.4718\pm0.0000$ & $0.9644\pm0.0000$ \\
\bottomrule
\end{tabular}
\end{table}

\clearpage
\begin{table}[!tp]
\centering
\caption{Perturbation-effect prediction: Norman, held-out double perturbations in K562. 39 intermediate double perturbations, grouped by effect size. Values are mean $\pm$ SD across five seed means, each averaged over five evaluation splits (Appendix~\ref{app:vc_metrics}).}
\label{tab:full:norman_combo:middle}
\small
\setlength{\tabcolsep}{3pt}
\begin{tabular}{@{}lcccc@{}}
\toprule
Model & Pearson $\Delta\uparrow$ & MAE $\downarrow$ & Overlap@N $\uparrow$ & \shortstack{L1 discrimination\\$\uparrow$} \\
\midrule
\textbf{Occam} & $0.8604\pm0.0009$ & $0.01254\pm0.00007$ & $0.2852\pm0.0061$ & $0.8975\pm0.0253$ \\
TxPert & $0.7838\pm0.0085$ & $0.01679\pm0.00020$ & $0.3442\pm0.0025$ & $0.8152\pm0.0110$ \\
STATE & $0.6681\pm0.0348$ & $0.01816\pm0.00067$ & $0.2812\pm0.0260$ & $0.7304\pm0.0140$ \\
CellFlow & $0.7147\pm0.0287$ & $0.01614\pm0.00022$ & $0.2656\pm0.0101$ & $0.7486\pm0.0309$ \\
PerturbDiff & $0.6052\pm0.1023$ & $0.02025\pm0.00259$ & $0.1755\pm0.0222$ & $0.6633\pm0.0320$ \\
Sum of singles & $0.8423\pm0.0006$ & $0.01455\pm0.00001$ & $0.2875\pm0.0005$ & $0.9409\pm0.0008$ \\
Mean perturbation & $0.4959\pm0.0012$ & $0.01862\pm0.00001$ & $0.1446\pm0.0005$ & $0.4818\pm0.0013$ \\
Control mean & $0.0079\pm0.0000$ & $0.01987\pm0.00000$ & $0.1353\pm0.0000$ & $0.4605\pm0.0000$ \\
Control cell & $0.0003\pm0.0038$ & $0.02146\pm0.00001$ & $0.0000\pm0.0000$ & $0.4622\pm0.0006$ \\
\midrule
Experimental reproducibility & $0.8917\pm0.0000$ & $0.01213\pm0.00000$ & $0.5281\pm0.0000$ & $0.9841\pm0.0000$ \\
\bottomrule
\end{tabular}
\end{table}

\begin{table}[!tp]
\centering
\caption{Perturbation-effect prediction: Norman, held-out double perturbations in K562. All 111 test double perturbations. Values are mean $\pm$ SD across five seed means, each averaged over five evaluation splits (Appendix~\ref{app:vc_metrics}).}
\label{tab:full:norman_combo:all}
\small
\setlength{\tabcolsep}{3pt}
\begin{tabular}{@{}lcccc@{}}
\toprule
Model & Pearson $\Delta\uparrow$ & MAE $\downarrow$ & Overlap@N $\uparrow$ & \shortstack{L1 discrimination\\$\uparrow$} \\
\midrule
\textbf{Occam} & $0.8526\pm0.0007$ & $0.01225\pm0.00004$ & $0.2806\pm0.0055$ & $0.8962\pm0.0117$ \\
TxPert & $0.7808\pm0.0072$ & $0.01653\pm0.00021$ & $0.3235\pm0.0018$ & $0.8346\pm0.0072$ \\
STATE & $0.6663\pm0.0335$ & $0.01770\pm0.00082$ & $0.2730\pm0.0238$ & $0.7939\pm0.0068$ \\
CellFlow & $0.7259\pm0.0203$ & $0.01556\pm0.00011$ & $0.2563\pm0.0090$ & $0.8163\pm0.0182$ \\
PerturbDiff & $0.6249\pm0.0871$ & $0.01957\pm0.00249$ & $0.1717\pm0.0213$ & $0.7182\pm0.0273$ \\
Sum of singles & $0.8333\pm0.0005$ & $0.01465\pm0.00001$ & $0.2831\pm0.0006$ & $0.9292\pm0.0011$ \\
Mean perturbation & $0.4549\pm0.0014$ & $0.01819\pm0.00002$ & $0.1341\pm0.0001$ & $0.5072\pm0.0017$ \\
Control mean & $0.0056\pm0.0000$ & $0.01894\pm0.00000$ & $0.1201\pm0.0000$ & $0.5045\pm0.0000$ \\
Control cell & $-0.0011\pm0.0043$ & $0.02065\pm0.00001$ & $0.0000\pm0.0000$ & $0.5058\pm0.0008$ \\
\midrule
Experimental reproducibility & $0.8822\pm0.0000$ & $0.01202\pm0.00000$ & $0.5087\pm0.0000$ & $0.9822\pm0.0000$ \\
\bottomrule
\end{tabular}
\end{table}

\clearpage
\begin{table}[!tp]
\centering
\caption{Perturbation-effect prediction: Norman, held-out double perturbations in K562. 37 strongly additive double perturbations, without effect-size grouping. Values are mean $\pm$ SD across five seed means, each averaged over five evaluation splits (Appendix~\ref{app:vc_metrics}).}
\label{tab:full:norman_combo:additive_raw}
\small
\setlength{\tabcolsep}{3pt}
\begin{tabular}{@{}lcccc@{}}
\toprule
Model & Pearson $\Delta\uparrow$ & MAE $\downarrow$ & Overlap@N $\uparrow$ & \shortstack{L1 discrimination\\$\uparrow$} \\
\midrule
\textbf{Occam} & $0.9050\pm0.0031$ & $0.01139\pm0.00020$ & $0.3046\pm0.0074$ & $0.9551\pm0.0078$ \\
TxPert & $0.8596\pm0.0083$ & $0.01425\pm0.00018$ & $0.3577\pm0.0019$ & $0.9586\pm0.0051$ \\
STATE & $0.7230\pm0.0362$ & $0.01689\pm0.00077$ & $0.3029\pm0.0317$ & $0.8638\pm0.0196$ \\
CellFlow & $0.7983\pm0.0153$ & $0.01453\pm0.00018$ & $0.2870\pm0.0090$ & $0.8936\pm0.0278$ \\
PerturbDiff & $0.6784\pm0.0977$ & $0.01933\pm0.00273$ & $0.1838\pm0.0247$ & $0.7131\pm0.0430$ \\
Sum of singles & $0.8936\pm0.0004$ & $0.01314\pm0.00002$ & $0.3053\pm0.0011$ & $0.9862\pm0.0003$ \\
Mean perturbation & $0.4893\pm0.0030$ & $0.01821\pm0.00004$ & $0.1484\pm0.0010$ & $0.4692\pm0.0036$ \\
Control mean & $0.0058\pm0.0000$ & $0.01945\pm0.00000$ & $0.1316\pm0.0000$ & $0.4670\pm0.0000$ \\
Control cell & $-0.0002\pm0.0101$ & $0.02098\pm0.00005$ & $0.0000\pm0.0000$ & $0.4661\pm0.0014$ \\
\midrule
Experimental reproducibility & $0.9261\pm0.0000$ & $0.01144\pm0.00000$ & $0.5473\pm0.0000$ & $0.9972\pm0.0000$ \\
\bottomrule
\end{tabular}
\end{table}

\begin{table}[!tp]
\centering
\caption{Perturbation-effect prediction: Norman, held-out double perturbations in K562. 37 weakly additive double perturbations, without effect-size grouping. Values are mean $\pm$ SD across five seed means, each averaged over five evaluation splits (Appendix~\ref{app:vc_metrics}).}
\label{tab:full:norman_combo:nonadditive_raw}
\small
\setlength{\tabcolsep}{3pt}
\begin{tabular}{@{}lcccc@{}}
\toprule
Model & Pearson $\Delta\uparrow$ & MAE $\downarrow$ & Overlap@N $\uparrow$ & \shortstack{L1 discrimination\\$\uparrow$} \\
\midrule
\textbf{Occam} & $0.7869\pm0.0059$ & $0.01160\pm0.00020$ & $0.2503\pm0.0043$ & $0.9047\pm0.0032$ \\
TxPert & $0.6907\pm0.0094$ & $0.01807\pm0.00030$ & $0.2668\pm0.0023$ & $0.7903\pm0.0086$ \\
STATE & $0.5998\pm0.0295$ & $0.01715\pm0.00096$ & $0.2196\pm0.0171$ & $0.8673\pm0.0139$ \\
CellFlow & $0.6724\pm0.0170$ & $0.01464\pm0.00051$ & $0.2154\pm0.0075$ & $0.8795\pm0.0010$ \\
PerturbDiff & $0.5714\pm0.0825$ & $0.01767\pm0.00248$ & $0.1560\pm0.0204$ & $0.8466\pm0.0174$ \\
Sum of singles & $0.7673\pm0.0011$ & $0.01499\pm0.00003$ & $0.2701\pm0.0006$ & $0.9128\pm0.0012$ \\
Mean perturbation & $0.3614\pm0.0038$ & $0.01607\pm0.00004$ & $0.1121\pm0.0006$ & $0.6418\pm0.0019$ \\
Control mean & $0.0035\pm0.0000$ & $0.01567\pm0.00000$ & $0.0957\pm0.0000$ & $0.6586\pm0.0000$ \\
Control cell & $0.0005\pm0.0112$ & $0.01755\pm0.00005$ & $0.0000\pm0.0000$ & $0.6609\pm0.0013$ \\
\midrule
Experimental reproducibility & $0.8228\pm0.0000$ & $0.01108\pm0.00000$ & $0.4603\pm0.0000$ & $0.9724\pm0.0000$ \\
\bottomrule
\end{tabular}
\end{table}

\clearpage
\begin{table}[!tp]
\centering
\caption{Perturbation-effect prediction: Norman, held-out double perturbations in K562. 37 intermediate double perturbations, without effect-size grouping. Values are mean $\pm$ SD across five seed means, each averaged over five evaluation splits (Appendix~\ref{app:vc_metrics}).}
\label{tab:full:norman_combo:middle_raw}
\small
\setlength{\tabcolsep}{3pt}
\begin{tabular}{@{}lcccc@{}}
\toprule
Model & Pearson $\Delta\uparrow$ & MAE $\downarrow$ & Overlap@N $\uparrow$ & \shortstack{L1 discrimination\\$\uparrow$} \\
\midrule
\textbf{Occam} & $0.8660\pm0.0036$ & $0.01375\pm0.00014$ & $0.2869\pm0.0057$ & $0.8289\pm0.0280$ \\
TxPert & $0.7920\pm0.0067$ & $0.01728\pm0.00016$ & $0.3459\pm0.0030$ & $0.7550\pm0.0111$ \\
STATE & $0.6760\pm0.0354$ & $0.01907\pm0.00082$ & $0.2964\pm0.0255$ & $0.6507\pm0.0110$ \\
CellFlow & $0.7070\pm0.0293$ & $0.01751\pm0.00028$ & $0.2663\pm0.0106$ & $0.6757\pm0.0263$ \\
PerturbDiff & $0.6249\pm0.0833$ & $0.02171\pm0.00228$ & $0.1753\pm0.0198$ & $0.5949\pm0.0291$ \\
Sum of singles & $0.8391\pm0.0007$ & $0.01582\pm0.00002$ & $0.2740\pm0.0008$ & $0.8886\pm0.0025$ \\
Mean perturbation & $0.5139\pm0.0014$ & $0.02028\pm0.00002$ & $0.1418\pm0.0008$ & $0.4104\pm0.0011$ \\
Control mean & $0.0076\pm0.0000$ & $0.02170\pm0.00000$ & $0.1328\pm0.0000$ & $0.3879\pm0.0000$ \\
Control cell & $-0.0035\pm0.0045$ & $0.02342\pm0.00002$ & $0.0000\pm0.0000$ & $0.3903\pm0.0003$ \\
\midrule
Experimental reproducibility & $0.8978\pm0.0000$ & $0.01354\pm0.00000$ & $0.5186\pm0.0000$ & $0.9769\pm0.0000$ \\
\bottomrule
\end{tabular}
\end{table}

\clearpage

\section{Training compute and inference throughput}
\label{supp_speedup}
This section estimates training compute for cell-representation models and measures cell-encoder inference throughput.

\subsection{Estimating training compute}
\label{supp_flops_estimation}

We estimate training compute analytically from each model architecture and training configuration. We count each multiplication and addition as one floating-point operation (FLOP), following the convention used in neural scaling-law analyses~\citep{kaplan2020scalinglawsneurallanguage,hoffmann2022trainingcomputeoptimallargelanguage}. For a linear layer with input dimension $d_{\mathrm{in}}$ and output dimension $d_{\mathrm{out}}$, the forward cost is
\begin{equation}
    C_{\mathrm{linear}}
    =
    2 d_{\mathrm{in}} d_{\mathrm{out}}
\end{equation}
FLOPs per input vector. For token-based models, this cost is multiplied by the sequence length $S$ when the same projection is applied independently to each token. Thus, the FLOPs of input projections, MLP/feed-forward blocks, output projections, and decoder heads are obtained by summing the corresponding linear-layer costs over all layers.

For transformer-based models, we additionally include the quadratic self-attention cost. In a transformer layer with sequence length $S$ and hidden dimension $d_{\mathrm{model}}$, the attention matrix computation $QK^\top$ costs approximately $2S^2d_{\mathrm{model}}$ FLOPs per sample, and the attention-weighted value aggregation costs another $2S^2d_{\mathrm{model}}$ FLOPs. Softmax and normalization have lower computational cost than these matrix multiplications and are not included. The attention contribution is therefore estimated as
\begin{equation}
    C_{\mathrm{attn}}
    =
    4 S^2 d_{\mathrm{model}}
\end{equation}
FLOPs per transformer layer and per sample. The total forward cost per sample, denoted $C_{\mathrm{fwd}}$, is obtained by summing all linear, attention, and decoder-head contributions in the forward pass.

We approximate the backward pass as twice the forward-pass cost, accounting for gradients with respect to both activations and parameters. The total training cost is therefore
\begin{equation}
    C_{\mathrm{train}}
    =
    3 C_{\mathrm{fwd}} D,
\end{equation}
where $D$ is the total number of training examples processed over all epochs, i.e.
\begin{equation}
    D
    =
    n_{\mathrm{profiles}} \times n_{\mathrm{epochs}}.
\end{equation}
Using more GPUs can change training time, but does not change this estimated number of operations. For MLP-only models without attention, the same estimate reduces approximately to
\begin{equation}
    C_{\mathrm{train}}
    \approx
    6 P D,
\end{equation}
where $P$ is the number of trainable parameters involved in dense matrix multiplications. This follows because the forward pass of a dense MLP costs approximately $2P$ FLOPs per sample, and forward plus backward training costs approximately three times the forward pass.

\begin{table*}[!tp]
\centering
\caption{\textbf{Training compute and inference throughput.}
All models were trained on the same H100 GPU cluster on DiverseRNA. Training compute is reported as total training operations, in petaFLOPs. Inference throughput is measured on approximately 100,000 profiles on one H100 80GB GPU; entries report profiles per second, so higher is better. Batch size is varied to show throughput scaling, and omitted where the tested configuration does not fit in 80GB memory. The tested scGPT and STATE configurations use maximum sequence lengths of 1,200 and 2,048 gene tokens. The 44k-gene comparison includes Occam and TxFM. These encoder measurements exclude perturbation-effect prediction and pretraining of external gene and protein embeddings.}
\label{tab:compute_efficiency}

\begingroup
\scriptsize
\setlength{\tabcolsep}{3pt}
\renewcommand{\arraystretch}{1.05}

\resizebox{\textwidth}{!}{%
\begin{tabular}{l |r| ccc | ccc | ccc}
\toprule
\multirow{2}{*}{\textbf{Model}}
& \multirow{2}{*}{\textbf{Train FLOPs} $\downarrow$}
& \multicolumn{3}{c}{\textbf{Inf. samples/s 1,024 genes} $\uparrow$}
& \multicolumn{3}{c}{\textbf{Inf. samples/s 2,048 genes} $\uparrow$}
& \multicolumn{3}{c}{\textbf{Inf. samples/s 44k genes} $\uparrow$} \\
\cmidrule(lr){3-5}
\cmidrule(lr){6-8}
\cmidrule(lr){9-11}
& \textbf{(Peta FLOPs)}
& \textbf{B=512} & \textbf{B=1024} & \textbf{B=2048}
& \textbf{B=512} & \textbf{B=1024} & \textbf{B=2048}
& \textbf{B=32} & \textbf{B=1024} & \textbf{B=2048} \\
\midrule
scGPT-90M
& 33,203
& 1,933 & 1,955 & 1,665
& --  & --  & --
& --  & --  & -- \\

STATE-90M
& 916,752
& 1,276 & 1,284 & --
& 566 & -- & --
& --  & --  & -- \\

TxFM-B
& 427,050
& 827 & 860 & --
& 372 & -- & --
& 4 & -- & -- \\

\midrule
\textsc{Occam-90M}
& \textbf{327}
& \textbf{1,448,625} & \textbf{1,619,255} & \textbf{2,345,992}
& \textbf{1,448,625} & \textbf{1,619,255} & \textbf{2,345,992}
& \textbf{102,915} & \textbf{1,619,255} & \textbf{2,345,992} \\
\bottomrule
\end{tabular}%
}

\endgroup
\end{table*}

\subsection{Measuring encoder throughput}
\label{supp_inference_speed}

We measure how many expression profiles each encoder processes per second on one NVIDIA H100 80GB GPU. Synthetic inputs match each model's expected input format, number of genes or gene tokens, and sparsity. Inputs are already on the GPU when timing starts, so the measurement excludes data loading, preprocessing, and transfer to the GPU.

All measurements are performed in inference mode, with gradient computation disabled and mixed-precision computation enabled using \texttt{bfloat16}. Token-based models use the gene-token sequence length specified for each benchmark setting. MLP models retain the fixed gene vocabulary used during training. We test batch sizes $B\in\{512,1024,2048\}$ when GPU memory allows. For the full 44,728-gene setting, we report $B\in\{32,1024,2048\}$; 32 is the largest TxFM batch that fits on the tested GPU in this setting.

For each model and batch size, we use the following protocol:
\begin{enumerate}
    \item Allocate one synthetic batch of size $B$ directly on the GPU.
    \item Run $W=10$ warm-up forward passes, which are discarded, to stabilize CUDA kernel launch overhead, memory allocation effects, and autotuning.
    \item Run $N=\lfloor 100{,}000/B \rfloor$ timed forward passes using the same pre-allocated batch.
    \item Synchronize the GPU immediately before and after the timed loop so that the measurement includes all scheduled operations.
    \item Record the elapsed time $T$.
\end{enumerate}

Throughput is reported as the number of samples processed per second:
\begin{equation}
    \mathrm{throughput}
    =
    \frac{NB}{T}
    \quad [\mathrm{samples/s}].
\end{equation}

These measurements describe encoder computation only. They exclude the time needed to read data, preprocess expression, and transfer inputs to the GPU.

\section{Representation evaluation details}
\label{supp_evaluation_details}
This section describes cell-representation evaluation with frozen encoders and gene-representation evaluation with phenotype probes. Perturbation-effect prediction uses the separate split-half protocol in Appendix~\ref{supp_experimental_accuracy}.

\subsection{Evaluation datasets}
\label{app:repr_datasets}

\paragraph{Perturbation representation datasets.}

The perturbation representation benchmark of~\citet{bendidi2024benchmarkpca} is evaluated on three large-scale Perturb-seq datasets, each profiling single-gene CRISPR perturbations in a single cell line. Table~\ref{tab:pert_datasets} summarizes their key statistics.

\begin{table}[!tp]
\centering
\caption{Perturbation representation benchmark datasets. Each dataset profiles single-gene CRISPR perturbations in a single cell line. Non-targeting guide cells serve as controls.}
\label{tab:pert_datasets}
\small
\begin{tabular}{lccc}
\toprule
& \textbf{HepG2} & \textbf{Jurkat} & \textbf{RPE1} \\
& \citep{nadig} & \citep{nadig} & \citep{replogle} \\
\midrule
Cell line       & HepG2      & Jurkat     & RPE1 \\
Cells           & 145{,}473  & 262{,}956  & 247{,}914 \\
Genes           & 9{,}624    & 8{,}882    & 8{,}749 \\
Perturbations   & 2{,}394    & 2{,}394    & 2{,}394 \\
Control cells   & 4{,}976    & 12{,}013   & 11{,}485 \\
Perturbed cells & 140{,}497  & 250{,}943  & 236{,}429 \\
Batches         & 56         & 55         & 56 \\
\bottomrule
\end{tabular}
\end{table}

All three datasets provide target-gene perturbation labels, experimental batch identifiers, and a control indicator. Embedding extraction follows the preprocessing expected by each model; Occam uses total-count scaling to 2,000 and $\log(1+\cdot)$. The downstream benchmark then applies the embedding transformations described below.

We use seed 42 to divide each dataset into a probe-training set and two test sets. Probe training uses 70\% of batches and 70\% of perturbation labels. Both test sets use held-out batches: test-seen contains perturbations observed during probe training, while test-unseen contains held-out perturbations. Cells with held-out perturbations in training batches are excluded. These assignments apply to probe training and evaluation; the cell encoder remains frozen. Control cells follow the assignment of their batch.

\paragraph{Cell-type integration datasets.}

The scIB benchmark~\citep{scib} is evaluated on two annotated scRNA-seq atlas datasets with known cell-type labels and batch structure. Table~\ref{tab:scib_datasets} summarizes their statistics.

\begin{table}[!tp]
\centering
\caption{scIB benchmark datasets. Each dataset provides cell-type annotations and batch labels for assessing biological conservation and batch correction.}
\label{tab:scib_datasets}
\small
\begin{tabular}{lcc}
\toprule
& \textbf{Pancreas} & \textbf{PBMC-68k} \\
\midrule
Cells       & 16{,}382  & 68{,}579 \\
Genes       & 19{,}093  & 32{,}738 \\
Cell types  & 14        & 11 \\
Batches     & 6         & 8 \\
\bottomrule
\end{tabular}
\end{table}

\paragraph{Perturbation-effect prediction datasets.} Appendix~\ref{app:vc_cohorts} describes the RHJ, K562, H1, and Norman datasets, including which perturbations and cell lines are held out and which genes are scored. The representation benchmarks above assess information in observed-cell embeddings, separately from prediction of perturbed expression.

\subsection{Perturbation representation benchmark}
\label{app:perturbation_repr_metrics}

We evaluate six tasks on frozen cell embeddings from each of the three perturbation datasets. Tasks 1 and 2 use Typical Variance Normalization (TVN), task 3 uses embeddings centered and scaled using controls, and tasks 4--6 use the encoder outputs directly. Within each benchmark run, stochastic probes use seeds 42 and 43, and their scores are averaged. These probe seeds are separate from the five repetitions summarized in the main representation table.

\paragraph{Embedding post-processing.}

Raw embeddings are the encoder outputs. To center and scale them, we estimate the mean and standard deviation of each embedding dimension from control cells within each batch, then apply that transformation to all cells in the batch. TVN adds three steps: PCA fitted on controls, a second centering and scaling in PCA space, and CORAL whitening within each batch. The whitening transformation uses the inverse square root of the regularized control covariance matrix to adjust correlations between embedding dimensions.

\paragraph{Task 1: Gene relationship recall.}

We average cell embeddings within each perturbation and compute cosine similarities between the resulting perturbation embeddings. We then compare these similarities with known gene relationships in CORUM, HuMAP, Reactome, SIGNOR, and STRING. For each database, recall is the fraction of known gene pairs whose similarity falls in the extreme tails of the full similarity distribution. The score averages recall across the five databases and is scaled to $[0,100]$.

\paragraph{Task 2: Perturbation consistency.}

This task asks whether cells receiving the same perturbation have more similar embeddings than expected by chance. We compare each cell with the mean embedding of the other cells receiving that perturbation, using cosine similarity. Randomly sampled cells provide the comparison expected by chance. The score is the fraction of perturbations with significant consistency at $p\leq0.01$, scaled to $[0,100]$. Non-targeting controls are excluded from the final score.

\paragraph{Task 3: iLISI batch integration.}

For each perturbation with sufficient cells, iLISI measures batch mixing on the $k$-nearest-neighbor graph of cell embeddings. Higher values indicate that cells receiving the same perturbation have neighbors from several batches. The score averages over qualifying perturbations and is scaled to $[0,100]$.

\paragraph{Task 4: KNN separability.}

We train a weighted $k$-nearest-neighbor (KNN) classifier to predict perturbation labels from frozen cell embeddings. The test-seen set contains the same perturbation labels in held-out batches. The score averages top-1 and top-5 classification accuracy.

\paragraph{Task 5: Linear separability.}

We train a linear classifier to predict perturbation labels from frozen cell embeddings. Control cells are excluded from training and evaluation. The score averages top-1 and top-5 classification accuracy on the test-seen set.

\paragraph{Task 6: Linear invertibility.}

We train a decoder to reconstruct gene expression from frozen cell embeddings and evaluate it on held-out perturbations in held-out batches. The score averages two measures: per-gene Spearman correlation between reconstructed and measured expression, and structural transcriptome distance, which assesses preservation of expression structure after centering within batches. The combined score is scaled to $[0,100]$.

\paragraph{Aggregation.}

Each task produces a score in $[0,100]$. For each dataset, the perturbation representation score is the arithmetic mean of the six task scores. The overall perturbation representation benchmark score is the arithmetic mean of the three per-dataset scores.

\subsection{scIB cell-integration benchmark}
\label{supp_scIBBenchmark}

We compute ten metrics from the scIB framework~\citep{scib} on each of the two cell-type datasets. We construct a $k$-nearest-neighbor graph from the embeddings using Euclidean distance. Leiden clustering is performed at the resolution that maximizes agreement with reference labels.

\paragraph{Biological conservation metrics.}

The biological conservation score averages six metrics. ASW\textsubscript{label} measures cell-type separation using average silhouette width. Normalized mutual information (NMI) and adjusted Rand index (ARI) measure agreement between Leiden clusters and reference cell-type annotations. Isolated-label ASW and isolated-label F1 evaluate whether rare or batch-isolated cell types remain distinguishable. cLISI measures cell-type neighborhood purity.

\paragraph{Batch correction metrics.}

The batch correction score averages four metrics. ASW\textsubscript{batch} measures batch mixing within cell types. Graph connectivity measures whether cells of the same type remain connected across batches. iLISI measures batch mixing in the neighborhood graph. PCR\textsubscript{batch} measures the reduction in batch-associated variance explained by principal components.

\paragraph{Aggregation.}

The biological conservation score is the arithmetic mean of the six biological metrics, and the batch correction score is the arithmetic mean of the four batch metrics. The per-dataset scIB score is
\begin{equation}
\label{eq:scib_score_appendix}
    S_{\mathrm{scIB}}
    =
    0.6\,S_{\mathrm{bio}}
    +
    0.4\,S_{\mathrm{batch}},
\end{equation}
following the scIB weighting convention~\citep{scib}. The overall scIB score is $\tfrac12(S_{\mathrm{Pancreas}}+S_{\mathrm{PBMC\text{-}68k}})$. We compute it from the per-dataset scores reported in the appendix and round to two decimal places; both datasets have equal weight.

\subsection{Gene-representation evaluation with phenotype probes}
\label{supp_gene_phenotype_eval}

We evaluate gene embeddings using the phenotype probes of~\citet{genbiobenchmark}. Each probe is trained to predict a gene's knockdown phenotype from its embedding. The probe receives only the gene embedding, which may include information from control cells. This differs from the full perturbation-effect predictor, which receives a control cell and a perturbation. The gene benchmark uses five-fold cross-validation over target genes, separately from split-half evaluation of perturbation-effect prediction.

\paragraph{Experimental data.}

The benchmark uses CRISPR-based single-gene knockdown experiments across three human cell lines: RPE1, HepG2, and Jurkat. For each cell line, approximately $2{,}386$ genes are individually perturbed, and transcriptomic responses are measured with single-cell RNA sequencing. Non-targeting guide cells serve as the unperturbed control population for each cell line.

\paragraph{Log-fold-change targets.}

We normalize each cell to a total count of $L=10{,}000$, apply $\log(1+\cdot)$, and average the resulting profiles within each perturbation and cell line. For perturbation $p$ in cell line $c$, the target is the difference between mean perturbed and control log-expression:
\begin{equation}
    \Delta \mathbf{x}_p^{(c)}
    =
    \bar{\mathbf{x}}_p^{(c)}
    -
    \bar{\mathbf{x}}_{\mathrm{ctrl}}^{(c)}
    \in \mathbb{R}^{G},
\end{equation}
where $G=6{,}640$ is the number of measured genes. This target captures the direction and magnitude of the transcriptional response to perturbation $p$.

\paragraph{Differential expression targets.}

The benchmark also constructs a categorical target that labels each measured gene as down-regulated, unchanged, or up-regulated in response to a perturbation. Differential expression is computed by comparing perturbed cells against non-targeting control cells using a Wilcoxon rank-sum test, followed by Benjamini--Hochberg correction. Genes with adjusted $p$-value below $0.05$ are labeled as differentially expressed, and the sign of the test statistic determines the direction of regulation. This yields a label vector
$\mathbf{y}_p^{(c)}\in\{-1,0,+1\}^{G}$ for each perturbation and cell line.

\paragraph{Gene embeddings.}

For models with gene tokens, we encode non-targeting control cells and average each gene's token representation across those cells. This gives one embedding per gene in each cell line. Occam constructs its cell-line-specific gene embeddings as described in Appendix~\ref{supp_Occam_gene_embeddings}. Models with static gene or protein embeddings use those embeddings directly. If a perturbed gene has no valid embedding, the benchmark assigns its fallback prediction.

\paragraph{Phenotype probe tasks.}

For log-fold-change prediction, a regression probe maps the perturbed gene's embedding to the response vector $\Delta\mathbf x_p^{(c)}$ across all $G$ measured genes. The benchmark uses KNN regression and regularized linear regression after standardization and PCA. For differential-expression prediction, a classification probe predicts whether each measured gene is downregulated, unchanged, or upregulated.

\paragraph{Evaluation protocol.}

We evaluate each cell line separately using five-fold cross-validation over perturbed genes. For log-fold-change prediction, the primary metric is $L_2$ error between predicted and measured response vectors, averaged over held-out perturbations. Secondary metrics are MSE, MAE, Spearman correlation, and Pearson correlation. For differential-expression prediction, the primary metric is per-gene $F_1$, averaged over genes; secondary metrics are precision, recall, area under the receiver operating characteristic curve (ROC AUC), and average precision. These tasks assess how well gene embeddings support prediction of knockdown phenotypes.

\section{Detailed representation results}
\label{supp_results_breakdown}

This section reports the full metric-level breakdown of the cell-representation results in Table~\ref{tab:main_results} and the additional gene-embedding phenotype benchmark in Table~\ref{tab:supp_genbio_aggregate}. 
For compactness, metric names are abbreviated in the tables. Higher is better unless a downward arrow is shown.

\paragraph{Abbreviations.}
For the perturbation representation benchmark, Rel. denotes gene-relationship recall, Cons. denotes perturbation consistency, iLISI denotes batch integration within perturbation groups, KNN denotes nonlinear perturbation separability, Lin. denotes linear perturbation separability, and Inv. denotes linear invertibility back to expression. 
For scIB, ASW-L denotes label average silhouette width, Iso-ASW and Iso-F1 denote isolated-label metrics, ASW-B denotes batch average silhouette width, GC denotes graph connectivity, and PCR denotes principal-component regression batch correction. 
For the gene-embedding benchmark, LFC-$L_2$ denotes mean $L_2$ error on predicted log-fold-change vectors, and DEG-F1 denotes per-gene $F_1$ score for differential-expression class prediction.

\begin{table*}[!tp]
\centering
\caption{\textbf{Representation-learning benchmark aggregates.}
Trained methods use DiverseRNA~\citep{txfm}, with 1.4 million expression profiles; Random is untrained. For each model, we report the perturbation representation score on HepG2, Jurkat, and RPE1, and the scIB score on Pancreas and PBMC-68k. Avg. denotes the arithmetic mean across datasets within each benchmark. Simple baselines are highlighted in blue.}
\label{tab:supp_repr_aggregate}
\scriptsize
\setlength{\tabcolsep}{3pt}
\renewcommand{\arraystretch}{1.05}

\resizebox{\textwidth}{!}{
\begin{tabular}{l|ccc|c|cc|c}
\toprule
\textbf{Model}
& \multicolumn{4}{c}{\textbf{Perturbation representation} $\uparrow$}
& \multicolumn{3}{c}{\textbf{scIB} $\uparrow$} \\
\cmidrule(lr){2-5}
\cmidrule(lr){6-8}
& \textbf{RPE1} & \textbf{HepG2} & \textbf{Jurkat} & \textbf{Avg.}
& \textbf{Panc.} & \textbf{PBMC-68k} & \textbf{Avg.} \\
\midrule
\rowcolor{baselineblue}
Random & 19.37 & 19.28 & 19.15 & 19.27 & 43.59 & 47.71 & 45.65 \\
scGPT-90M & 24.25 & 24.31 & 23.47 & 24.01 & 37.39 & 53.66 & 45.53 \\
scMAE-90M & 26.50 & 25.85 & 24.77 & 25.71 & 47.12 & 51.70 & 49.41 \\
STATE-SE-90M & 35.86 & 32.14 & 30.78 & 32.93 & 47.03 & 48.38 & 47.71 \\
scVI-110M & 35.55 & 33.29 & 30.62 & 33.16 & 48.72 & 55.02 & 51.87 \\
\rowcolor{baselineblue}
Linear AE (22M) & 36.80 & 33.93 & 31.73 & 34.15 & 51.14 & 54.20 & 52.67 \\
\rowcolor{baselineblue}
PCA & 38.34 & 35.05 & 32.59 & 35.33 & 47.48 & 54.56 & 51.02 \\
\rowcolor{baselineblue}
Ridge-PCA & 38.54 & 35.18 & 32.82 & 35.51 & 47.46 & 54.42 & 50.94 \\
TxFM-B & 42.05 & 37.50 & 36.81 & 38.79 & 47.29 & 53.14 & 50.22 \\
\midrule
\textsc{Occam-90M} & 42.94 & 38.79 & 36.22 & 39.32 & 53.10 & 54.33 & 53.72 \\
\bottomrule
\end{tabular}
}
\end{table*}

\begin{table*}[!tp]
\centering
\caption{\textbf{Representation-learning benchmark aggregates.}
These public checkpoints use pretraining corpora larger than DiverseRNA. TrFormer-sp denotes TranscriptFormer-Sapiens. For each model, we report the perturbation representation score on HepG2, Jurkat, and RPE1, and the scIB score on Pancreas and PBMC-68k. Avg. denotes the arithmetic mean across datasets within each benchmark.}
\label{tab:supp_fm_aggregate}
\scriptsize
\setlength{\tabcolsep}{3pt}
\renewcommand{\arraystretch}{1.05}

\resizebox{\textwidth}{!}{
\begin{tabular}{l|ccc|c|cc|c}
\toprule
\textbf{Model}
& \multicolumn{4}{c}{\textbf{Perturbation representation} $\uparrow$}
& \multicolumn{3}{c}{\textbf{scIB} $\uparrow$} \\
\cmidrule(lr){2-5}
\cmidrule(lr){6-8}
& \textbf{RPE1} & \textbf{HepG2} & \textbf{Jurkat} & \textbf{Avg.}
& \textbf{Panc.} & \textbf{PBMC-68k} & \textbf{Avg.} \\
\midrule
scGPT-43M & 25.20 & 25.01 & 24.06 & 24.75 & 41.89 & 55.43 & 48.66 \\
TrFormer-sp & 31.50 & 28.57 & 27.11 & 29.06 & 52.22 & 49.17 & 50.70 \\
AIDO.Cell-100M & 34.56 & 32.21 & 30.90 & 32.56 & 44.95 & 53.23 & 49.09 \\
STATE-SE-600M & 39.54 & 34.97 & 33.19 & 35.90 & 46.58 & 50.83 & 48.71 \\
\bottomrule
\end{tabular}
}
\end{table*}

\begin{table*}[!tp]
\centering
\caption{\textbf{Gene-representation evaluation with phenotype probes.}
Supervised probes predict knockdown phenotypes from gene embeddings to evaluate representation quality. LFC-$L_2$ is the mean $L_2$ error for predicted log-fold-change vectors (lower is better). DEG-F1 is the per-gene $F_1$ score for differential-expression class prediction (higher is better). Avg. denotes the arithmetic mean across RPE1, HepG2, and Jurkat. Values are point estimates. TrFormer-sp denotes TranscriptFormer-Sapiens. Baselines are highlighted in blue; rows within each group are ordered by average LFC error.}
\label{tab:supp_genbio_aggregate}
\scriptsize
\setlength{\tabcolsep}{4pt}
\renewcommand{\arraystretch}{1.05}

\resizebox{\textwidth}{!}{
\begin{tabular}{l|ccc|c|ccc|c}
\toprule
\textbf{Model}
& \multicolumn{4}{c}{\textbf{LFC-$L_2$} $\downarrow$}
& \multicolumn{4}{c}{\textbf{DEG-F1} $\uparrow$} \\
\cmidrule(lr){2-5}
\cmidrule(lr){6-9}
& \textbf{RPE1} & \textbf{HepG2} & \textbf{Jurkat} & \textbf{Avg.}
& \textbf{RPE1} & \textbf{HepG2} & \textbf{Jurkat} & \textbf{Avg.} \\
\midrule
\multicolumn{9}{l}{\textit{Pretrained gene representations}} \\
\textsc{Occam-90M} & 7.37 & 6.84 & 6.42 & 6.87 & 0.381 & 0.434 & 0.385 & 0.400 \\
STATE-SE-600M        & 7.40 & 6.85 & 6.43 & 6.89 & 0.383 & 0.426 & 0.388 & 0.399 \\
AIDO.Cell-100M         & 7.41 & 6.87 & 6.43 & 6.91 & 0.366 & 0.410 & 0.373 & 0.383 \\
ESM2         & 7.42 & 6.91 & 6.45 & 6.93 & 0.361 & 0.398 & 0.369 & 0.376 \\
scGPT-43M        & 7.50 & 6.90 & 6.45 & 6.95 & 0.369 & 0.407 & 0.377 & 0.384 \\
TrFormer-sp & 7.67 & 7.06 & 6.51 & 7.08 & 0.332 & 0.356 & 0.343 & 0.344 \\
\midrule
\multicolumn{9}{l}{\textit{Simple baselines}} \\
\rowcolor{baselineblue}
PCA          & 7.62 & 7.00 & 6.49 & 7.03 & 0.353 & 0.406 & 0.364 & 0.375 \\
\rowcolor{baselineblue}
Random       & 7.65 & 7.04 & 6.50 & 7.06 & 0.324 & 0.344 & 0.350 & 0.340 \\
\bottomrule
\end{tabular}
}
\end{table*}

\begin{table*}[!tp]
\centering
\caption{\textbf{Perturbation representation benchmark on RPE1.}
We report all six task scores used to compute the RPE1 perturbation representation average. Rel. measures recovery of known gene relationships; Cons. measures within-perturbation consistency; iLISI measures batch mixing among cells receiving the same perturbation; KNN and Lin. measure perturbation-label classification with KNN and linear probes; Inv. measures reconstruction of gene expression from the embeddings. Avg. is the arithmetic mean of the six tasks. Higher is better for all metrics. Simple baselines are highlighted in blue; public denotes a public checkpoint.}
\label{tab:supp_pert_repr_rpe1}
\scriptsize
\setlength{\tabcolsep}{4pt}
\renewcommand{\arraystretch}{1.05}

\begin{tabular}{l|cccccc|c}
\toprule
\textbf{Model} & \textbf{Rel.} & \textbf{Cons.} & \textbf{iLISI} & \textbf{KNN} & \textbf{Lin.} & \textbf{Inv.} & \textbf{Avg.} \\
\midrule
\rowcolor{baselineblue}
Random        & 10.31 & 0.00 & 70.34 & 8.18 & 2.49 & 24.88 & 19.37 \\ 
scGPT-90M (DiverseRNA)   & 25.55 & 3.13 & 71.13 & 9.23 & 4.14 & 32.31 & 24.25 \\ 
scGPT-43M (public)    & 30.13 & 3.55 & 70.14 & 9.21 & 5.22 & 32.93 & 25.20 \\ 
scMAE-90M  (DiverseRNA)    & 26.36 & 2.83 & 71.00 & 10.25 & 10.99 & 37.57 & 26.50 \\ 
TrFormer-sp (public) & 36.12 & 31.30 & 55.83 & 11.46 & 15.72 & 38.55 & 31.50 \\  
AIDO.Cell-100M  (public)   & 43.21 & 36.19 & 69.11 & 12.31 & 9.06 & 37.46 & 34.56 \\
scVI-110M  (DiverseRNA)        & 39.16 & 28.42 & 70.21 & 13.34 & 19.34 & 42.87 & 35.55 \\  
STATE-SE-90M (DiverseRNA)         & 41.79 & 30.00 & 69.97 & 13.41 & 21.75 & 38.23 & 35.86 \\ 
\rowcolor{baselineblue}
Linear AE (22M)  (DiverseRNA)    & 43.45 & 28.50 & 71.09 & 13.95 & 22.83 & 40.94 & 36.80 \\  
\rowcolor{baselineblue}
PCA  (DiverseRNA)   & 44.67 & 32.93 & 70.63 & 15.15 & 25.91 & 40.77 & 38.34 \\   
\rowcolor{baselineblue}
Ridge-PCA (DiverseRNA)  & 44.87 & 32.89 & 70.65 & 15.08 & 26.87 & 40.88 & 38.54 \\  
STATE-SE-600M (public)         & 44.43 & 40.83 & 69.10 & 13.61 & 24.22 & 45.02 & 39.54 \\ 
TxFM-B (DiverseRNA)          & 46.12 & 44.59 & 65.98 & 16.94 & 37.23 & 41.45 & 42.05 \\ 
\midrule
\textsc{Occam-90M} (DiverseRNA) & 46.15 & 44.46 & 70.22 & 16.82 & 36.70 & 43.27 & 42.94 \\
\bottomrule
\end{tabular}
\end{table*}
\begin{table*}[!tp]
\centering
\caption{\textbf{Perturbation representation benchmark on HepG2.}
We report all six task scores used to compute the HepG2 perturbation representation average. Rel. measures recovery of known gene relationships; Cons. measures within-perturbation consistency; iLISI measures batch mixing among cells receiving the same perturbation; KNN and Lin. measure perturbation-label classification with KNN and linear probes; Inv. measures reconstruction of gene expression from the embeddings. Avg. is the arithmetic mean of the six tasks. Higher is better for all metrics. Simple baselines are highlighted in blue; public denotes a public checkpoint.}
\label{tab:supp_pert_repr_hepg2}
\scriptsize
\setlength{\tabcolsep}{4pt}
\renewcommand{\arraystretch}{1.05}

\begin{tabular}{l|cccccc|c}
\toprule
\textbf{Model} & \textbf{Rel.} & \textbf{Cons.} & \textbf{iLISI} & \textbf{KNN} & \textbf{Lin.} & \textbf{Inv.} & \textbf{Avg.} \\
\midrule
\rowcolor{baselineblue}
Random        & 9.64 & 0.00 & 70.33 & 5.44 & 1.13 & 29.16 & 19.28 \\  
scGPT-90M (DiverseRNA)   & 29.74 & 1.59 & 70.89 & 6.40 & 2.37 & 34.86 & 24.31 \\ 
scGPT-43M (public)     & 31.53 & 2.38 & 69.65 & 6.50 & 3.35 & 36.62 & 25.01 \\   
scMAE-90M  (DiverseRNA)     & 26.93 & 0.38 & 71.00 & 8.05 & 7.36 & 41.43 & 25.85 \\
TrFormer-sp (public) & 35.24 & 19.39 & 53.50 & 8.48 & 12.13 & 42.69 & 28.57 \\  
STATE-SE-90M (DiverseRNA)         & 40.66 & 17.51 & 69.83 & 9.57 & 12.95 & 42.32 & 32.14 \\      
AIDO.Cell-100M  (public)   & 42.60 & 27.00 & 68.49 & 8.54 & 5.64 & 40.99 & 32.21 \\
scVI-110M  (DiverseRNA)        & 42.19 & 18.35 & 69.94 & 10.56 & 13.08 & 45.61 & 33.29 \\  
\rowcolor{baselineblue}
Linear AE (22M)  (DiverseRNA)    & 44.22 & 18.93 & 70.89 & 10.31 & 15.08 & 44.12 & 33.93 \\ 
STATE-SE-600M (public)         & 44.55 & 24.66 & 68.45 & 10.29 & 15.83 & 46.07 & 34.97 \\
\rowcolor{baselineblue}
PCA  (DiverseRNA) & 45.51 & 21.27 & 70.41 & 10.95 & 17.63 & 44.51 & 35.05 \\  
\rowcolor{baselineblue}
Ridge-PCA (DiverseRNA)  & 45.43 & 21.27 & 70.42 & 10.94 & 18.37 & 44.63 & 35.18 \\  
TxFM-B (DiverseRNA)          & 45.37 & 33.05 & 65.61 & 11.63 & 24.98 & 44.33 & 37.50 \\
\midrule
\textsc{Occam-90M} (DiverseRNA) & 47.67 & 32.09 & 69.90 & 11.50 & 25.47 & 46.08 & 38.79 \\   
\bottomrule
\end{tabular}
\end{table*}
\begin{table*}[!tp]
\centering
\caption{\textbf{Perturbation representation benchmark on Jurkat.}
We report all six task scores used to compute the Jurkat perturbation representation average. Rel. measures recovery of known gene relationships; Cons. measures within-perturbation consistency; iLISI measures batch mixing among cells receiving the same perturbation; KNN and Lin. measure perturbation-label classification with KNN and linear probes; Inv. measures reconstruction of gene expression from the embeddings. Avg. is the arithmetic mean of the six tasks. Higher is better for all metrics. Simple baselines are highlighted in blue; public denotes a public checkpoint.}
\label{tab:supp_pert_repr_jurkat}
\scriptsize
\setlength{\tabcolsep}{4pt}
\renewcommand{\arraystretch}{1.05}

\begin{tabular}{l|cccccc|c}
\toprule
\textbf{Model} & \textbf{Rel.} & \textbf{Cons.} & \textbf{iLISI} & \textbf{KNN} & \textbf{Lin.} & \textbf{Inv.} & \textbf{Avg.} \\
\midrule
\rowcolor{baselineblue}
Random        & 10.58 & 0.00 & 70.63 & 7.44 & 1.29 & 24.98 & 19.15 \\ 
scGPT-90M (DiverseRNA)    & 26.83 & 0.33 & 71.12 & 7.93 & 3.03 & 31.55 & 23.47 \\   
scGPT-43M (public)   & 30.05 & 0.33 & 69.93 & 8.02 & 3.67 & 32.35 & 24.06 \\
scMAE-90M  (DiverseRNA)      & 25.82 & 3.25 & 70.00 & 9.31 & 4.49 & 35.79 & 24.77 \\
TrFormer-sp (public) & 32.70 & 17.85 & 56.67 & 8.94 & 9.57 & 36.91 & 27.11 \\
scVI-110M  (DiverseRNA)     & 37.60 & 15.38 & 70.22 & 10.05 & 9.43 & 41.06 & 30.62 \\         
STATE-SE-90M (DiverseRNA)         & 38.14 & 17.85 & 69.86 & 10.21 & 12.15 & 36.45 & 30.78 \\ 
AIDO.Cell-100M  (public)  & 38.88 & 25.71 & 69.58 & 9.46 & 5.89 & 35.89 & 30.90 \\ 
\rowcolor{baselineblue}
Linear AE (22M)  (DiverseRNA)   & 39.41 & 16.68 & 71.05 & 10.37 & 13.37 & 39.47 & 31.73 \\
\rowcolor{baselineblue}
PCA  (DiverseRNA) & 40.83 & 19.86 & 70.56 & 11.04 & 14.11 & 39.17 & 32.59 \\
\rowcolor{baselineblue}
Ridge-PCA (DiverseRNA)  & 41.04 & 19.82 & 70.54 & 11.01 & 15.26 & 39.30 & 32.82 \\  
STATE-SE-600M (public)         & 39.22 & 25.17 & 68.40 & 10.64 & 11.95 & 43.79 & 33.19 \\
TxFM-B (DiverseRNA)          & 42.54 & 33.49 & 67.91 & 12.51 & 23.80 & 40.60 & 36.81 \\
\midrule
\textsc{Occam-90M} (DiverseRNA) & 41.72 & 30.23 & 70.37 & 11.97 & 21.86 & 41.21 & 36.22 \\  
\bottomrule
\end{tabular}
\end{table*}

\begin{table*}[!tp]
\centering
\caption{\textbf{scIB benchmark on Pancreas.}
We report all scIB component metrics. Biological conservation metrics are ASW-L, NMI, ARI, Iso-ASW, Iso-F1, and cLISI; their mean is Bio. Batch correction metrics are ASW-B, graph connectivity (GC), iLISI, and PCR; their mean is Batch. The final scIB score is $0.6\,\mathrm{Bio}+0.4\,\mathrm{Batch}$. Higher is better for all metrics. Simple baselines are highlighted in blue; public denotes a public checkpoint.}
\label{tab:supp_scib_pancreas}
\scriptsize
\setlength{\tabcolsep}{3pt}
\renewcommand{\arraystretch}{1.05}

\resizebox{\textwidth}{!}{
\begin{tabular}{l|cccccc|c|cccc|c|c}
\toprule
\textbf{Model}
& \multicolumn{7}{c|}{\textbf{Biological conservation}}
& \multicolumn{5}{c|}{\textbf{Batch correction}}
& \textbf{Final} \\
\cmidrule(lr){2-8}
\cmidrule(lr){9-13}
\cmidrule(lr){14-14}
& \textbf{ASW-L}
& \textbf{NMI}
& \textbf{ARI}
& \textbf{Iso-ASW}
& \textbf{Iso-F1}
& \textbf{cLISI}
& \textbf{Bio}
& \textbf{ASW-B}
& \textbf{GC}
& \textbf{iLISI}
& \textbf{PCR}
& \textbf{Batch}
& \textbf{scIB} \\
\midrule
scGPT-90M (DiverseRNA)    & 34.44 & 25.73 & 10.08 & 28.18 & 10.00 & 90.31 & 33.12 & 47.26 & 46.54 & 1.25 & 80.14 & 43.80 & 37.39 \\  
scGPT-43M (public)   & 41.61 & 35.27 & 15.28 & 41.16 & 5.41 & 94.80 & 38.92 & 65.28 & 51.19 & 0.60 & 68.34 & 46.35 & 41.89 \\
\rowcolor{baselineblue}
Random        & 48.28 & 0.49 & 0.0 & 48.24 & 3.48 & 78.07 & 29.76 & 95.39 & 26.65 & 35.39 & 99.93 & 64.34 & 43.59 \\
AIDO.Cell-100M  (public)  & 45.35 & 49.69 & 22.21 & 56.97 & 8.42 & 88.51 & 45.19 & 72.62 & 86.94 & 2.15 & 16.67 & 44.59 & 44.95 \\ 
STATE-SE-600M (public)         & 51.56 & 61.83 & 26.23 & 51.60 & 2.47 & 93.99 & 47.95 & 94.33 & 83.80 & 0.00 & 0.00 & 44.53 & 46.58 \\ 
STATE-SE-90M (DiverseRNA)  & 51.61 & 49.62 & 21.08 & 66.01 & 11.20 & 87.49 & 47.84 & 83.20 & 86.94 & 0.00 & 13.15 & 45.82 & 47.03 \\    
scMAE-90M  (DiverseRNA)      & 50.48 & 8.99 & 3.80 & 50.74 & 15.48 & 93.37 & 37.14 & 93.89 & 34.65 & 33.39 & 86.43 & 62.09 & 47.12 \\
TxFM-B (DiverseRNA)           & 50.91 & 50.68 & 17.90 & 65.61 & 13.19 & 99.49 & 49.63 & 84.29 & 87.73 & 0.00 & 3.06 & 43.77 & 47.29 \\
\rowcolor{baselineblue}
Ridge-PCA (DiverseRNA) & 50.81 & 49.02 & 20.38 & 63.73 & 35.90 & 98.77 & 53.10 & 77.85 & 78.13 & 0.00 & 0.00 & 39.00 & 47.46 \\
\rowcolor{baselineblue}
PCA  (DiverseRNA) & 50.82 & 48.59 & 20.48 & 63.78 & 35.90 & 98.86 & 53.07 & 77.84 & 78.51 & 0.00 & 0.00 & 39.09 & 47.48 \\
scVI-110M  (DiverseRNA)    & 51.38 & 49.92 & 20.66 & 59.90 & 26.42 & 98.81 & 51.18 & 87.65 & 85.69 & 0.00 & 6.76 & 45.03 & 48.72 \\
\rowcolor{baselineblue}
Linear AE (22M)  (DiverseRNA)   & 51.86 & 52.69 & 32.69 & 62.54 & 36.84 & 98.74 & 55.89 & 83.63 & 84.02 & 0.00 & 8.35 & 44.00 & 51.14 \\   
TrFormer-sp (public) & 48.12 & 57.26 & 39.12 & 74.19 & 10.94 & 88.44 & 53.01 & 75.79 & 85.60 & 42.76 & 0.00 & 51.04 & 52.22 \\
\midrule
\textsc{Occam-90M} (DiverseRNA) & 53.95 & 65.65 & 40.34 & 52.96 & 9.45 & 99.88 & 53.70 & 90.48 & 95.73 & 0.00 & 22.61 & 52.20 & 53.10 \\  
\bottomrule
\end{tabular}
}
\end{table*}
\begin{table*}[!tp]
\centering
\caption{\textbf{scIB benchmark on PBMC-68k.}
We report all scIB component metrics. Biological conservation metrics are ASW-L, NMI, ARI, Iso-ASW, Iso-F1, and cLISI; their mean is Bio. Batch correction metrics are ASW-B, graph connectivity (GC), iLISI, and PCR; their mean is Batch. The final scIB score is $0.6\,\mathrm{Bio}+0.4\,\mathrm{Batch}$. Higher is better for all metrics. Simple baselines are highlighted in blue; public denotes a public checkpoint.}
\label{tab:supp_scib_PBMC-68k}
\scriptsize
\setlength{\tabcolsep}{3pt}
\renewcommand{\arraystretch}{1.05}

\resizebox{\textwidth}{!}{
\begin{tabular}{l|cccccc|c|cccc|c|c}
\toprule
\textbf{Model}
& \multicolumn{7}{c|}{\textbf{Biological conservation}}
& \multicolumn{5}{c|}{\textbf{Batch correction}}
& \textbf{Final} \\
\cmidrule(lr){2-8}
\cmidrule(lr){9-13}
\cmidrule(lr){14-14}
& \textbf{ASW-L}
& \textbf{NMI}
& \textbf{ARI}
& \textbf{Iso-ASW}
& \textbf{Iso-F1}
& \textbf{cLISI}
& \textbf{Bio}
& \textbf{ASW-B}
& \textbf{GC}
& \textbf{iLISI}
& \textbf{PCR}
& \textbf{Batch}
& \textbf{scIB} \\
\midrule
\rowcolor{baselineblue}
Random        & 49.90 & 0.16 & 0.01 & 49.87 & 15.43 & 69.95 & 30.88 & 99.10 & 52.65 & 52.98 & 87.06 & 72.95 & 47.71 \\    
STATE-SE-90M (DiverseRNA)   & 50.23 & 45.76 & 22.09 & 40.46 & 35.18 & 72.80 & 44.42 & 97.12 & 80.46 & 39.76 & 0.00 & 54.33 & 48.38 \\ 
TrFormer-sp (public) & 49.89 & 50.11 & 20.48 & 50.75 & 17.61 & 72.53 & 43.56 & 94.72 & 81.95 & 53.66 & 0.00 & 57.58 & 49.17 \\ 
STATE-SE-600M (public)         & 50.09 & 45.61 & 23.29 & 45.26 & 41.38 & 78.60 & 47.37 & 99.53 & 78.19 & 46.36 & 0.00 & 56.02 & 50.83 \\  
scMAE-90M  (DiverseRNA)     & 50.70 & 5.16 & 2.51 & 50.87 & 24.43 & 97.35 & 38.50 & 98.30 & 58.65 & 49.98 & 79.06 & 71.50 & 51.70 \\  
TxFM-B (DiverseRNA)          & 50.06 & 46.92 & 24.68 & 49.46 & 47.18 & 83.80 & 50.35 & 95.75 & 81.42 & 52.14 & 0.00 & 57.33 & 53.14 \\
AIDO.Cell-100M  (public)  & 47.66 & 45.34 & 18.37 & 46.90 & 50.90 & 89.50 & 49.78 & 93.71 & 81.75 & 58.19 & 0.00 & 58.41 & 53.23 \\        
scGPT-90M (DiverseRNA)         & 48.20 & 49.00 & 22.60 & 48.30 & 49.10 & 88.20 & 50.90 & 93.80 & 77.60 & 59.79 & 0.00 & 57.80 & 53.66 \\
\rowcolor{baselineblue}
Linear AE (22M)  (DiverseRNA) & 49.65 & 45.72 & 23.04 & 49.65 & 48.48 & 87.05 & 50.60 & 97.74 & 80.06 & 60.62 & 0.00 & 59.61 & 54.20 \\
\rowcolor{baselineblue}
Ridge-PCA (DiverseRNA) & 49.89 & 46.07 & 22.21 & 49.89 & 51.88 & 89.15 & 51.52 & 97.34 & 79.26 & 58.53 & 0.00 & 58.78 & 54.42 \\ 
\rowcolor{baselineblue}
PCA  (DiverseRNA) & 49.90 & 47.72 & 23.13 & 49.89 & 49.95 & 89.14 & 51.62 & 97.36 & 79.40 & 59.15 & 0.00 & 58.98 & 54.56 \\  
scVI-110M  (DiverseRNA)   & 50.34 & 49.01 & 23.82 & 50.55 & 52.11 & 88.52 & 52.39 & 98.70 & 79.33 & 57.80 & 0.00 & 58.96 & 55.02 \\
scGPT-43M (public)    & 50.07 & 50.97 & 24.54 & 50.22 & 51.09 & 90.50 & 52.90 & 95.62 & 79.68 & 61.65 & 0.00 & 59.24 & 55.43 \\
\midrule
\textsc{Occam-90M} (DiverseRNA) & 49.74 & 49.09 & 23.08 & 50.35 & 51.32 & 84.83 & 51.40 & 98.45 & 82.68 & 53.76 & 0.00 & 58.73 & 54.33 \\  
\bottomrule
\end{tabular}
}
\end{table*}

\end{document}